\documentclass[a4paper,12pt]{article}
\usepackage[T1]{fontenc}
\usepackage[utf8]{inputenc}
\usepackage{lmodern}
\usepackage{graphicx}
\usepackage{epstopdf} 
\usepackage{color}
\usepackage{bbold}
\usepackage{amsmath}
\title{Pseudoscalar meson mass relations from a second-order phase transition}

\author{R. L. P. G. Amaral$^{a}$\footnote{email: rubensamaral@id.uff.br} ,
	V. E. R. Lemes$^{b}$\footnote{email: verlemes@gmail.com},
	O. S. Ventura$^{c}$\footnote{email: ozemar.ventura@cefet-rj.br}, L.C.Q.Vilar $^{b}$\footnote{email: lcqvilar@gmail.com}  \\
	\small \em $^a$Instituto de F\'{\i}sica, Universidade Federal do Fluminense\\
	\small \em Av. Litor\^anea S/N, Boa Viagem, Niter\'oi-RJ CEP. 24210-340,
	Brazil\\
	\small \em $^b$Instituto de F\'\i sica, Universidade do Estado do Rio de
	Janeiro,\\
	\small \em Rua S\~{a}o Francisco Xavier 524, Maracan\~{a}, Rio de Janeiro - RJ,
	20550-013, Brazil\\
	\small \em $^c$Departamento de F\'\i sica, Centro Federal de Educa\c{c}\~ao Tecnol\'ogica do Rio de
	Janeiro,\\
	\small\em Av.Maracan\~a 249, 20271-110, Rio de Janeiro - RJ, Brazil}

\begin{document}
	\maketitle
	\begin{abstract}	
This work is divided in two parts. The first three sections review meson physics phenomenology, highlighting the history of pseudoscalar multiplet mass spectra research. We then propose a new approach for the mass mixing problem based exclusively on a second order phase transition principle. This development leads to new relations among the masses of the mesons in this nonet, which present a nice agreement with measurements. The quark constitutions and the mixing angle problem are also correctly addressed. After describing the spontaneous symmetry breaking process coming from this phenomenological analysis, we establish a field theory with the necessary elements that could reproduce theoretically such results. This is done in the second part of this work, where the one-loop quantum corrections are computed in full, with the cubic and quartic scalar vertices treated on the same footing. The resulting mass Lagrangian has a universal flavor structure, which forces a refined identification between the scalar components and the $\eta^{0}$--$\eta^{\prime\,0}$ states; we derive the general matching condition and present its minimal solution, which reproduces the phenomenological mixing matrix exactly. Combined with the gap equation of the scalar sector, the model is left with a single free coupling, which does not affect the mass predictions. We emphasize that this is a phenomenological model constructed to reproduce certain mass relations, not a first-principles derivation from QCD. Nevertheless, as we will show, the model's predictions for the pseudoscalar meson masses agree remarkably well with the observed QCD spectrum.
		
\end{abstract}
\section{Introduction}

The pseudoscalar meson spectrum poses a long-standing challenge to effective descriptions of strong-interaction physics. While Quantum Chromodynamics (QCD) is accepted as the fundamental theory of the strong force, the derivation of low-energy observables---such as the masses and mixing of the $\eta$ and $\eta'$ mesons---from first principles remains a formidable task. Consequently, phenomenological models that capture essential symmetry-breaking patterns and reproduce empirical mass relations continue to provide valuable insight, particularly when they point to structural features that might otherwise remain hidden in the full theory.

In this work we adopt such a perspective. We construct an effective field theory of scalar and fermionic fields possessing a chiral $SU(3)\times SU(3)$ symmetry that is spontaneously broken via a second-order phase transition. The model, which is \emph{not} QCD, incorporates effective fermionic and scalar degrees of freedom, analogous to the light quarks and the mesonic fields, respectively. Our goal is not to derive the pseudoscalar masses from the QCD Lagrangian, but to identify a minimal symmetry-breaking pattern that reproduces the observed mass relations, quark content, and mixing angle of the $0^-$ nonet with high precision. This pattern involves both the octet and the singlet directions in a way that lies outside the standard Gell-Mann--Okubo hypothesis, and it emerges naturally in a complex gauge-theoretic framework previously developed in connection with the Gribov confinement problem \cite{ALVV20}. This suggests that the model may capture structural aspects of the QCD vacuum even though it is not obtained from QCD itself.

To motivate the specific breaking pattern we propose, it is useful to recall the historical development of meson phenomenology.
 Since the discovery of the pions \cite{lattes}, organizing mesons into multiplets according to an approximate higher symmetry became a central objective. The pioneering work of Nambu and Goldstone \cite{namprl, nampr, gold1961} established spontaneous symmetry breaking (SSB) as the natural explanation for the pion's role as a nearly massless Goldstone boson. The relevant symmetry was soon identified as $SU(2)\times SU(2)$ \cite{adler65, weis65}, and the recognition that this chiral symmetry is a natural property of a fermionic Dirac Lagrangian
\begin{eqnarray}
\cal{L} &=& \overline{q}\gamma^{\mu} \left\{D_{\mu}+ v_{\mu} (x) + \gamma_{5} a_{\mu} (x) \right\} q - \overline{q} \left\{ s (x) -\gamma_{5} p(x) \right\} q 
\label{action}
\end{eqnarray}
shifted the theoretical emphasis toward a Lagrangian description \cite{wein68}. Here $q$ and $\overline{q}$ denote fermionic fields that, in the context of QCD, are associated with confined quarks and anti-quarks, while $v_{\mu}$, $a_{\mu}$, $s$ and $p$ are hermitian external fields encoding chiral currents.
$D_{\mu}$ stands for the covariant derivative that includes the gluonic fields. 

In order to describe the first three observed mesons $\pi^{+}$, $\pi^{-}$ and $\pi^{0}$, an isospin structure was idealized for the quark field, whose two flavor states were then called up and down quarks.
\begin{equation}
q = \left( \begin{array}{c} u \\ d\\
\end{array} \right) .
\label{q}
\end{equation}
Under independent right and left transformations
\begin{equation}
q^{'}(x) =\left\{ \frac{1}{2} \left( 1  + \gamma_{5} \right) V_{R}(x) +  \frac{1}{2} \left( 1  - \gamma_{5} \right) V_{L}(x) \right\} q(x) ,
\label{deltaq}
\end{equation}
in an infinitesimal form
\begin{eqnarray}
V_{R}(x) &=& 1 + i \alpha (x) + i \beta  (x) , \nonumber \\
V_{L}(x)  &=& 1 + i \alpha (x) - i \beta  (x) ,
\label{V}
\end{eqnarray}
the Lagrangian $(\ref{action})$ becomes invariant if the external fields get transformed under the induced "gauge transformations" \cite{Gasser1983,Gasser1984}

\begin{eqnarray}
\delta v_{\mu} &=& \partial_{\mu} \alpha  + i [\alpha ,  v_{\mu} ] + i [ \beta ,   a_{\mu} ] , \nonumber \\
\delta a_{\mu}  &=&  \partial_{\mu}  \beta   + i [\alpha ,  a_{\mu} ] + i [ \beta ,   v_{\mu} ] ,  \nonumber \\
\delta s &=& i [\alpha ,  s ] - \{ \beta ,  p \}  ,  \nonumber \\
\delta p &=& i [\alpha ,  p ] + \{ \beta ,  s \}.
\label{deltav}
\end{eqnarray}

In the actual world, quarks are not massless as the Lagrangian $(\ref{action})$  might suggest, but this can be accommodated if we suppose that the external field $s(x)$ acquires a non-null vacuum expectation value by means of a SSB mechanism not fully understood. In this way, the $ SU(2 $)x$SU(2) $ symmetry expressed in  $(\ref{deltaq})$, 
$(\ref{V})$ and $(\ref{deltav})$ gets spontaneously broken to a vectorial $SU(2)$ symmetry. The idea is that such breaking would lead to the generation of three massless Goldstone bosons, the pions. Describing the broken phase as an effective field theory allows a glimpse of how pions could be interpreted as quark condensates \cite{weinbook}.

This picture had to change fast at the beginning of the 1960's. Four new pseudoscalar mesons, the kaons, joined the pions. There were now seven $0^{-}$ light mesons (at the same time there were also eight $ \frac{1}{2} ^{+}$ barions cataloged). The isospin $ SU(2) $ group became too small for this plethora of particles. The early Sakata model \cite{saka56} anticipated an $ SU(3)$ structure, but as the quark model was not yet available, his attempt to describe mesons as been constituted from barions was not fruitful. It took some years for Gell-Mann \cite{gell61} and Ne'eman \cite{neem61} to independently propose what became known as the Eightfold-Way. If we join another "light" meson $0^{-}$, the $\eta$ particle, we can understand these eight mesons as an octet, an irreducible representation of the $SU(3)$ symmetry. The transformations associated to this symmetry should act on a triplet of basic particles, demanding another light quark in the Lagrangian $(\ref{action})$, the strange quark $s$

\begin{equation}
q = \left( \begin{array}{c} u \\ d\\ s\\
\end{array} \right) .
\label{triq}
\end{equation}
In this way, the chiral  $ SU(2 $)x$SU(2) $ was extended to an  $ SU(3 $)x$SU(3) $ symmetry, preserving the same structure of $(\ref{action})$,  $(\ref{deltaq})$, 
$(\ref{V})$ and $(\ref{deltav})$ with $(\ref{triq})$ replacing $(\ref{q})$. Gell-Mann sustained that this symmetry must be spontaneously broken to a diagonal $SU(3) $ subgroup, and we reinforce again that this hypothesis has no field theoretical description up to now. But there is a novelty in this new $SU(3) $ picture. Although pions have similar masses implying an approximate diagonal $SU(2) $  symmetry, kaons and $\eta$ are particles rather heavier than the pions, meaning that the residual $SU(3) $ symmetry should itself be badly broken. This mass difference could be explained by the fact that such new mesons would be made up of at least one $s$ quark, heavier than $u$ or $d$ from which pions are built. Anyway, such symmetry in the broken phase would seem to be useless \cite{gibpol}. But Gell-Mann thought otherwise. He wrote the complete basis of 3x3 matrices of $SU(3)$ generators, the Gell-Mann basis, and noticed that there was a diagonal generator that commutes with the isospin $SU(2)$ subgroup. His hypothesis was that the breaking of the residual $SU(3)$ was aligned in the direction of this generator in the internal space. In this case, this breaking would preserve the isospin symmetry, and consequently the approximate degeneracy of the pions masses. Furthermore, this simple idea allowed him to predict experimentally testable consequences. These were called the  Gell-Mann-Okubo (GMO) mass relations \cite{gmo}. These are formulae that relate masses of mesons in the same octet (the same occurs for barions in octets or decouplets).

We will now summarize the development of the GMO relations, paying attention to the historical problems of the meson octets. The reader should keep in mind, however, that the phenomenological relations to be derived in the following sections are not claimed to be exact predictions of QCD, but rather properties of a specific field-theoretic model that exhibits a second-order phase transition. Whether the resulting mass relations closely resemble those observed in QCD will be examined in the course of the paper.

\section{The GMO Mesons Mass Relations}
Before presenting our own framework, it is useful to review the standard phenomenological description of the pseudoscalar octet within QCD-inspired effective field theory. This summary serves two purposes: it fixes the notation and the experimental benchmarks that any successful model must meet, and it highlights the specific tensions in the established approach---particularly in the $\eta$--$\eta'$ sector---that motivate the alternative construction developed in the next sections. The reasoning below follows the effective field theory perspective of Weinberg \cite{weinbook}, which remains the closest available QCD-based formulation of the spontaneous symmetry breaking (SSB) process for the chiral $SU(3)\times SU(3)$ symmetry.

The first hypothesis present in \cite{weinbook} is that the mesons of the pseudoscalar $0^{-}$ octet should be accommodated in an adjoint matricial 3x3 representation 

\begin{equation}
 B=\begin{bmatrix}
	\frac{1}{\sqrt{2}}\pi^{0} + \frac{1}{\sqrt{6}}\eta^0
 & \pi^{+} & K^{+}\\
	 \pi^{-} & -\frac{1}{\sqrt{2}}\pi^{0} + \frac{1}{\sqrt{6}}\eta^0	 &  K^{0}\\
		\bar{K}^{-} & 	\bar{K}^{0} & -\frac{2}{\sqrt{6}}\eta^0 \\
	\end{bmatrix}  \, .
\label{b}
\end{equation}
This realization appears already in the original construction of Ne'eman  \cite{neem61}. It can be empirically justified for the need of the coupling of such fields to the external currents of ($\ref{action}$) \cite{Boul1982}. This argument certainly shows how advantageous this arrangement is, but it is not a theoretical requirement, and there is no known explanation of why mesons should be organized this way.

In $(\ref{b})$ we can understand that each particle is associated to a given generator of the $SU(3)$ Gell-Mann basis (or independent combinations). They are thought as Goldstone fields after the chiral SSB. The effective field theory approach then prescribes that these Goldstone modes should be isolated from the other scalar modes through a field dependent gauge transformation, where the Goldstone fields appear as the gauge parameters. In the Lagrangian 
$(\ref{action})$, the effect of this transformation takes us to the definition of the "Goldstone free quark fields" $\tilde{q}$ \cite{weinbook}

\begin{equation}
q = exp \big(-iB \big) \tilde{q} .
\label{qtil}
\end{equation}

Now, as already mentioned, the external field $s(x)$ in $(\ref{action})$ must acquire a non-null vacuum expectation value $M_q$ to take account of the quark masses. This massive element after $(\ref{qtil})$ is redefined as
\begin{eqnarray}
{\cal{L}}_{mass} &=& - \overline{q} M_q q = - \bar{\tilde{q}} exp \big(iB \big) M_q exp \big(-iB \big) \tilde{q} ,
\label{mqaction}
\end{eqnarray}
with 
$$M_q =\begin{bmatrix}
	m_u & 0 & 0\\
	0 & m_d & 0\\
		0 & 	0 & m_s\\ 
	\end{bmatrix}$$ 
where $m_u , m_d , m_s$ stand for the masses of the quarks u, d and s respectively. Within this description, we can find the Gell-Mann hypothesis of the breaking of the residual $SU(3)$. Once we notice that experimentally $ m_u \approx m_d $ and $m_s >> m_u$, we see that this element explicitly breaks the $SU(3)$. In Gell-Mann's vision, this breaking was caused by an element transforming as the diagonal element of his $SU(3)$ basis preserving the $SU(2)$ subspace, explaining the degeneracy of the up and down quarks, and consequently of the pions. After some considerations regarding the vacuum expectation values of the quark bilinears in $(\ref{mqaction})$, the mesons masses can be read from this Lagrangian being related to the quark masses. Simple manipulations lead us to the GMO mass relations for the pseudoscalar octet

\begin{equation}
m_{\eta^0}^2=\frac{4}{3}m_K^2-\frac{1}{3}m_{\pi}^2 ,
\label{GMO}
\end{equation}
where $m_K$ and $m_\pi $ stands for the (average) mass of the kaons and pions respectively, and $m_{\eta^0}$ for the ${\eta^0}$ mass. This field theory approach is well suited to address these mass relations in the mesonic case, and can be extended in an obvious way for any other mesonic octet. Other approaches, based on group theory \cite{Itz80} or perturbation theory \cite{gibpol}, for example,  can address also the analogous formulae for the barionic octets and decuplets. The barionic $\frac{3}{2} ^+$ case deserves special mention as the discovery of the missing $\Omega ^-$ particle of this decuplet \cite{bar64} in $1964$, with the right quantum numbers and mass as predicted by Gell-Mann two years before, convinced physicists of the correctness of the Eightfold-Way with the $SU(3)$ breaking hypothesis.

The success of the GMO mass relations is well established for barions. Another example is the $\frac{1}{2} ^+$ octet,  where the neutron and proton are allocated, with an accuracy of 1 per cent of the mass formula \cite{gibpol}. But we cannot say the same in the mesonic case. When applied to the $0^-$ multiplet, equation $(\ref{GMO})$ predicts ~565 MeV for the $\eta^0$ mass, while experimentally this mass is around 548 MeV \cite{weinbook} (actually, in the case of other mesonic octets this problem gets worse \cite{gibpol}). This discrepancy is sufficient to state that this whole picture needs to be amended. The proposed solution that we have until today  is to adapt this scheme by saying that for such particles a ninth meson should be involved in the derivation of the mass formula. This new particle should be a singlet state, that would originally be degenerated in mass with each meson octet. This degeneracy would be responsible for the mixing of the masses, and in the end each multiplet of mesons becomes a nonet \cite{cca2002}. This argument is phenomenologically well based, although it is rather heuristic in the sense that no theoretical origin is given for this mixing. This works well for the majority of the mesonic nonets, but in the case of the pseudoscalar (and possibly scalar) multiplet, this adaptation still does not provide an acceptable prediction in face of the experimental data. This is the root of the famous $U(1)$ problem, that was eventually solved when instantons were taken into consideration. Nowadays, this story is receiving a plot twist coming from numerical simulations of lattice QCD.  As our main interest is the construction of a field theoretical basis for the chiral SSB, all these informations will be used to guide us, and so we will present them with some detail. 

As it happens in all mesonic octets, there is in fact a candidate for the ninth meson to join the $0^-$ octet. We will call it  the $\eta^{\prime\, 0}$ state (the actual particles $\eta$ and $\eta^{\prime}$ are those that would be obtained from such states after diagonalization of the mass matrix). In principle, this state was understood as the Goldstone meson resulting from a SSB of the global symmetry $U(1)$ x $U(1)$ of the Lagrangian $(\ref{action})$ \cite{weinbook} .  In this way, it must enter associated to the 3x3 identity generator  in the matricial representation $(\ref{b})$

\begin{equation}
 B^\prime=\begin{bmatrix}
	\frac{1}{\sqrt{2}}\pi^{0} + \frac{1}{\sqrt{6}}\eta^0 +\frac{1}{\sqrt{3}}\eta^{{\prime}{0}}
 & \pi^{+} & K^{+}\\
	 \pi^{-} & -\frac{1}{\sqrt{2}}\pi^{0} + \frac{1}{\sqrt{6}}\eta^0 +\frac{1}{\sqrt{3}}\eta^{{\prime}{0}}	 &  K^{0}\\
		\bar{K}^{-} & 	\bar{K}^{0} & -{{\frac{\sqrt 2}{\sqrt 3}}}\eta^0 +\frac{1}{\sqrt{3}}\eta^{{\prime}{0}} \\
	\end{bmatrix}  \, .
\label{bp}
\end{equation}

Then, before the mixing of these states, we establish the following constitution for the mesons in the diagonal as

\begin{equation}
\pi^{0} = \frac{1}{\sqrt{2}}[u\bar{u}-d\bar{d}] ,
\label{pi0}
\end{equation}

\begin{equation} {\eta^{0}}=\frac{1}{\sqrt{6}}(u\bar{u}+d\bar{d}-2s\bar{s})  ,
\label{eta0quarks}
\end{equation}

\begin{equation} {\eta}^{\prime 0}=\frac{1}{\sqrt{3}}(u\bar{u}+d\bar{d}+s\bar{s})  .
\label{etap0quarks}
\end{equation}

Following the same manipulations that before led us to the GMO relation $(\ref{GMO})$, we find a mass matrix for $\eta^{0}$  and $\eta^{\prime\, 0}$ 

\begin{equation}
M^2=\frac{1}{3}\begin{bmatrix}
	4m_K^2-m_{\pi}^2
	&-2\sqrt{2}(m_K^2-m_{\pi}^2)\\
	-2\sqrt{2}(m_K^2-m_{\pi}^2) & 2m_K^2+m_{\pi}^2	\end{bmatrix}
\label{m2}
\end{equation}

Here it is important to emphasize, in the context of the effective field theory approach, that this mixing makes sense if the origin of the  $\eta^{\prime\, 0}$ is again from a SSB process, as it is for the rest of the nonet. This is the reason that allows $\eta^{\prime\, 0}$ to enter the "Goldstone free quark fields" redefinition in $(\ref{qtil})$
which ultimately leads to the mixing mass matrix. Next we will see  how this observation becomes relevant.

The diagonalization of the mass matrix indicates a state, that we generically call ${\omega_0}$,  with a mass equal to

\begin{equation} 
m_{\omega_0}^2 = 2m_{K}^2 -m_{\pi}^2  ,
\label{mw0}
\end{equation}
actually even further than the GMO prediction coming from  $(\ref{GMO})$ for the $\eta$ mass. The other state ${\omega_0^\prime}$ has a predicted  mass from the diagonalization just equal to $m_{\pi}$, 
\begin{equation} 
m_{\omega_0^\prime}^2 = m_{\pi}^2  ,
\label{mwp0}
\end{equation}
which makes no sense, as the observed mass of the  $\eta^{\prime}$ is 958 Mev, whereas the pions have typical masses around 137 Mev. This is the $U(1)$ problem that we mentioned earlier \cite{weinbook}. If we look at the eigenvectors coming from this diagonalization, we find the constituents in terms of the quarks  $(\ref{triq})$ respectively as
 
 \begin{equation}
 \omega_0=s\bar{s} , 
 \label{w0}
 \end{equation}

\begin{equation}
{\omega_0^\prime} = \frac{1}{\sqrt{2}}[u\bar{u}+d\bar{d}] .
\label{wol}
\end{equation}

These constituents are also distinct from what is observed for the  $\eta$ and $\eta^{\prime}$ mesons \cite{gil1987, chau1990}

\begin{equation}
{\eta}=\frac{1}{\sqrt{3}}(u\bar{u}+d\bar{d}-s\bar{s})  ,
\label{etaquarks}
\end{equation}

\begin{equation}
{\eta}^{\prime}=\frac{1}{\sqrt{6}}(u\bar{u}+d\bar{d}+2s\bar{s})  .
\label{etapquarks}
\end{equation}
This diagonalized structure of the states ${\omega_0}$ and  ${\omega_0^\prime}$ is called the ideal mixing of the octet-singlet.  Although this ideal mixing picture does not work for the pseudoscalar (and scalar) mesons, which is the main problem that we want to study, we must here call attention to the fact that this picture functions nicely for the other mesons multiplets. In order to make this analysis easier, we will rename the fields in the previous formulae in order to fit any meson multiplet. Then $ m_{\pi}^2$ becomes $ m_{1}^2$, the mass of the isovector in any meson octet,  $m_{K}^2= m_{1/2}^2$ the mass of the isodoublet, $m_{\omega_0}^2 =m_{0}^2 $ the mass of the isoscalar mostly octet, and $m_{\omega_0^\prime}^2 =m_{0^\prime}^2 $ the mass of the isoscalar mostly singlet. Using such redefinitions in $(\ref{mw0})$ and $(\ref{mwp0})$ we obtain the GMO mass relation for an ideal octet-singlet mixing \cite{bura, bura1997, bura1998} 

\begin{equation} 
\frac{m_{1}^2 + m_{0^\prime}^2}{2} + m_{0}^2 = 2 m_{1/2}^2    .
\label{mGMO}
\end{equation}

We can test this formula for the vector $1^-$ 

\begin{table}[htbp]
\centering
\begin{tabular}{|l|c|c|}
\hline
 & PARTICLE & MASS (MeV) \\
\hline
isovector & $\rho$ &  $ m_{1}=775$  \\
mostly singlet & $\omega$ & $m_{0^\prime}=783$ \\
mostly octet & $\Phi$ & $m_{0}=1019$ \\
isodoublet & $K^*$ &  $ m_{1/2}=892$  \\
\hline
\end{tabular}
\caption{Vector meson masses.}
\label{tab:vector_mesons}
\end{table}

or the tensor $2^+$

\begin{table}[htbp]
\centering
\begin{tabular}{|l|c|c|}
\hline
 & PARTICLE & MASS (MeV) \\
\hline
isovector & $a_2$ &  $ m_{1}=1317$  \\
mostly singlet & $f_2$ & $m_{0^\prime}=1275$ \\
mostly octet & $f^{\prime}_2$ & $m_{0}=1517$ \\
isodoublet & $K^*_2$ &  $ m_{1/2}=1427$  \\
\hline
\end{tabular}
\caption{Tensor meson masses.}
\label{tab:tensor_mesons}
\end{table}

\noindent mesonic multiplets  (masses obtained from \cite{PDG2020}). First we should stress that, using this data for the $1^-$ multiplet,  the original GMO formula $(\ref{GMO})$ predicts the existence of a isoscalar with a mass around 930 MeV \cite{gibpol}. Both $\omega$ and $\Phi$ masses are very distant from this. The prediction of $(\ref{mGMO})$ for the isodoublet masses lies between 1 and 2 percent  of the observed masses. This shows how well the nonet concept works for mesons. We can notice that for such nonets  $1^-$  and  $2^+$ the masses of the isovector and isoscalar mostly singlet are very close, thus satisfying the condition  $(\ref{mwp0})$ necessary for an ideal mixing. There is no $U(1)$ problem for these multiplets. The reason is that in these nonets the ninth member is not a would-be Goldstone boson: it is a conventional quark-antiquark state whose mass is comparable to the octet members. Only in the pseudoscalar (and possibly scalar) case is the would-be ninth Goldstone boson removed by the axial anomaly: the instanton-induced interaction explicitly breaks the $U(1)_{A}$ symmetry and shifts the singlet mass before mixing, upward in the pseudoscalar channel (and with opposite sign in the scalar one).

It is important also to present the so called $\theta$ angle of the mix, a parameter frequently mentioned in the phenomenological literature. We can understand the diagonalization procedure of the mass matrix $M^2$ as a rotation from the states prior to the mixing in $(\ref{eta0quarks})$ and $(\ref{etap0quarks})$ to the eigenvectors $(\ref{w0})$ and $(\ref{wol})$ as

\begin{eqnarray}
{\omega_0} &=& \eta^0\cos\theta - \eta^{\prime\, 0}\sin\theta \, , \nonumber \\
{\omega_0}^{\prime} &=& \eta^0\sin\theta + \eta^{\prime\, 0}\cos\theta \, ,
\label{rotation}
\end{eqnarray}
and in this case we immediately obtain the ideal mixing angle $\theta_{id} = 35,3^{\circ}$. By simple manipulations, we can as well relate $\theta$ with the masses $ m_0^2$ and $ m_{0^\prime}^2$  of the states after the rotation and the predicted GMO mass
$m_{\eta_0}^2$ for the state $\eta_0$ before the mix
\begin{equation}
\tan^2\theta=\frac{m_{\eta^0}^2-m_0^2}{m_{0^\prime}^2-m_{\eta^0}^2}  .
\label{tgtheta}
\end{equation}
If in this expression we enter as $ m_0^2$ and  $ m_{0^\prime}^2$ the actually measured masses of the isoscalar particles, and calculate $m_{\eta^0}^2$ from the GMO prediction $(\ref{GMO})$ adapted for a general octet, we interpret $(\ref{tgtheta})$ as a formula for an experimental mixing angle for each nonet.

Then, for example, if we use the observed masses listed in Tables~\ref{tab:vector_mesons} and~\ref{tab:tensor_mesons} for the $1^-$ multiplet in Eq.~(\ref{tgtheta}), we obtain an experimental angle $\theta_{1^-} = 36^{\circ}$, showing how this nonet is close to the ideal mixing case. If we use the mass data of the pseudoscalar $0^-$ nonet, we find an angle $\theta_{0^-} = 10.6^{\circ}$. Such a small angle in Eq.~(\ref{rotation}) would imply that the quark content of the $\eta$ and $\eta^\prime$ particles should be rather close to those of the unmixed states $\eta^0$ and $\eta^{\prime\,0}$. However, this is not what we find from the measurements in Eqs.~(\ref{etaquarks}) and~(\ref{etapquarks}). The observed composition of these particles is compatible with an experimental angle near $-20^{\circ}$, which is obtained from several different phenomenological arguments \cite{gil1987,chau1990,Ball1995,Dighe1995,Cao2012}. If we substitute Eqs.~(\ref{etaquarks}) and~(\ref{etapquarks}) in place of $\omega_0$ and $\omega_0^\prime$ in Eq.~(\ref{rotation}), respectively, we obtain $\theta = \arctan (-1/(2\sqrt{2}))$, i.e. $\theta \approx -19.5^{\circ}$. Here we should highlight the prediction made by Bramon in 1974 of this exact angle based on duality arguments \cite{Bramon1974}.

All this together means that the concept of a nonet of mesons is a successful idea, but the picture of the GMO ideal mixing collapses in face of the results for the pseudoscalar multiplet. Put in a straightforward way, we can see the problem as a mass gap that appears in the diagonalization of the mass matrix $(\ref{m2})$. The trace of the matrix should be preserved, but if we take the observed masses of $\eta$ and $\eta^\prime$ that should be the result of such diagonalization we find the following condition for this trace \cite{Feldmann1999}

\begin{equation}
m_\eta^2+ m_{\eta^\prime}^2 =2m_K^2 + 3A  .
\label{mgap}
\end{equation} 
With the measured Kaon mass $m_K\approx494$  MeV, the mass shift $3A$ is about 0,73 Ge$V^2$. This is a measure of the $U(1)$ breaking. This scenario needed modifications.

The way for this modification was pointed out by 't Hooft \cite{tHooft1976}, suggesting an inter-quark interaction induced by instantons. The logic of this proposal follows this reasoning \cite{Munz1993}: if we imagine mesons being formed by condensates of quarks through a confining potential generated by an "one gluon exchange" (OGE), and knowing that this process leads to a flavor independent interaction, the imediate consequence is that the mass of the isovector state $(\ref{pi0})$  and that of the isoscalar mostly singlet $(\ref{wol})$ becomes degenarate as they have the same quark content. And as we have shown, this is in total contradiction with the experimental masses. Then, it seems to be necessary to search for a new interaction among quarks that circumvents this OGE obstacle, but at the same time only acting on the pseudoscalar (and scalar) multiplet. This is what is expected from the instantons \cite{Munz1993}. 

Instantons are topological solutions of the Yang-Mills theories in the Euclidean space. This way, they are not continually deformable into the perturbative gluon solutions, and then lead to effective quark interactions that are not covered by the perturbative approach. Furthermore, these interactions break the $U(1)_A$ chiral symmetry, that we have shown as the essence behind the degeneracy of the ideal mix masses. They appear as a consequence of the $U(1)$ Adler-Bell-Jackiw anomaly \cite{Adler1969,Bell1969}.

The idea is that this new quark interaction be responsible for the increase of the  $\eta^{\prime\, 0}$ before the mix. The observed mass shift $3A$ in $(\ref{mgap})$ would be caused by instantons \cite{Dmitrasinovic1996,Dmitrasinovic1997}. In this regard, the origin of the meson $\eta^{\prime\, 0}$ would not be as a Goldstone boson anymore \cite{tHooft1986}. This point of view makes perfect sense when we become aware of the quantum $U(1)_A$ anomaly. But there all still some points that remain obscure. For example, we understand the necessary mixing process in the $0^-$ multiplet as an effect of a ninth Goldstone boson joining the other eight in the matrix $(\ref{bp})$, and as we mentioned after $(\ref{etap0quarks})$, this was allowed by the SSB origin of  $\eta^{\prime\, 0}$. If $\eta^{\prime\, 0}$ is an instanton, and not a Goldstone boson, how can we understand the whole effective theory approach \cite{weinbook}? If we imagine the formation of the mesons as pairs quark-anti-quark at the moment of the SSB breaking as shown in $(\ref{pi0})$ to  $(\ref{etap0quarks})$, how can this be reconciled with the previous existence of $\eta^{\prime\, 0}$ as an instanton? Moreover, as we argued, the other mesonic multiplets also demand a ninth element in order that the GMO mass relations can be successfully developed after the ideal mix. In these cases, we do not have the intervention of the instanton, but then if the emergence of the ninth element cannot be assigned to the axial anomaly, and if at the same time we cannot presume the SSB of the $U(1)$ x $U(1)$ symmetry, how can the presence of these ninth elements be justified if they can be neither instantons nor Golsdstone bosons?

There is another difficulty for this intantonic scenario coming from the phenomenology of the problem, even if we ignore these theoretical questions. We will present this approach now. The proposal is that the instanton would affect the mass of  $\eta^{\prime\, 0}$ as an increase in the mixed mass matrix  $(\ref{m2})$ \cite{bura}

\begin{equation}
M^2=\frac{1}{3}\begin{bmatrix}
	4m_K^2-m_{\pi}^2
	&-2\sqrt{2}(m_K^2-m_{\pi}^2)\\
	-2\sqrt{2}(m_K^2-m_{\pi}^2) & 2m_K^2+m_{\pi}^2 +9A	\end{bmatrix}  \, \, \,    .
\label{9a}
\end{equation}
The first result is to satisfy the mass gap $(\ref{mgap})$ in an obvious way. With this adaptation, it is possible to arrive at a mixing angle near to the observed  $-20^{\circ}$  \cite{bura}. But it is also obvious that this is not sufficient, and we can see that we get a predicted mass for the $\eta$ coming from the diagonalization of $(\ref{9a})$ far from the actual observed mass \cite{bura1998}. This led the authors of this proposal to state that instantons, that may shift the mass of the pseudoscalar isoscalar singlet, cannot constitute the explanation for the failure of the mass formulae for the pseudoscalar nonet \cite{bura1998}. This point of view was then criticized with the argument that $(\ref{9a})$ has a built-in limiting hypothesis, that the transition between quarkonia is flavor independent \cite{Li2002}. This refers to an older generalization of the mixed mass matrix for the pseudoscalar nonet, even before the 
consideration of instanton effects \cite{Suura1974} (this development was derived independently some years later in \cite{Kawai1983}). In these works they show that even the original ideal mixing $(\ref{m2})$ has this hidden hypothesis that the transitions between a pair $q_i {\bar{q}}_i $, where i denotes flavor, to a different $q_j {\bar{q}}_j $ are not allowed. This condition can be relaxed. The conventional way to present the most general mixing matrix still based on the other general hypothesis of Gell-Mann leading to the ideal mix is just to use the ideal state vectors $(\ref{w0})$ and $(\ref{wol})$ as the initial basis for the mix, usually referred as stragionium and nonstrangionium states respectively. It is also called the proper states basis. In this basis the most general mix, including possible instantonic contributions, is described by

\begin{equation}
M_G^2 =\begin{bmatrix}
	m_{\omega_0^\prime}^2+2\alpha 
	&\sqrt{2}\alpha Z\\
	\sqrt{2}\alpha Z & m_{\omega_0}^2+\alpha Z^2 \end{bmatrix}
\label{mg2}
\end{equation}
where $m_{\omega_0^\prime}^2$ and $ m_{\omega_0}^2$ are the squared masses of the nonstrangionium and strangionium states, as given by the Gell-Mann's prescriptions $(\ref{mw0})$ and $(\ref{mwp0})$. If we use the rotation between the proper basis $(\omega_0^\prime,\omega_0 )$ and the original pre-mix basis $(\eta_0,\eta_0^\prime )$ obtained from $(\ref{eta0quarks})$ and $(\ref{etap0quarks})$

\begin{equation}
(\omega_0^\prime,\omega_0 )=(\eta,\eta_{0}^{\prime})R=(\eta,\eta_{0}^{\prime})\begin{bmatrix}
	\frac{1 }{\sqrt{3}}
	&-\sqrt{\frac{2 }{{3}}}\\
	\sqrt{\frac{2 }{{3}}} & \frac{1 }{\sqrt{3}} \end{bmatrix}  \, ,
\label{R}
\end{equation}
it is immediate to show that $(\ref{9a})$ is a particular case of 
$(\ref{mg2})$ with $\alpha = A$ and $ Z = 1 $. This last assumption in fact means that  the transition between quarkonia is flavor independent \cite{Li2002}. Thus, it is possible to generalize the mixing matrix by taking together with the instanton effect $A$, a general $Z$ in order to admit flavor dependent transitions, and $(\ref{mg2})$ is this most general case compatible with Gell-Mann's prescriptions. Actually,  \cite{Suura1974} and \cite{Kawai1983} allowed also for a possible mixing with a third pure glueball state, which is the subject of studies upon nowadays for several mesonic multiplets (see for example \cite{Klempt2021}), but there is not much room for such admixture in the $0^-$ nonet \cite{Rosner1982}. As the pseudoscalar case is our main interest, we ignore any possible glueball mixing. So, we proceed by describing a generic arbitrary mesonic state $\lvert{\psi}\rangle$ expanded in terms of the nonstrangionium state $(\ref{wol})$, now called $\lvert{N}\rangle $, and of the strangionium   $(\ref{w0})$, $\lvert{S}\rangle $, as

\begin{equation}
\lvert{\psi}\rangle=x\lvert{N}\rangle+y\lvert{S}\rangle \, ,
\label{psi}
\end{equation}
with
\begin{equation}
x^2+y^2=1  \, .
\label{1}
\end{equation}
We can write an eigenvector equation for the generic mixing matrix $(\ref{mg2})$ as 
\begin{equation}
M_G^2  \lvert{\psi}\rangle=m^{2}\lvert{\psi}\rangle \, ,
\label{eigen}
\end{equation}
and then obtain a relation that is independent of $\alpha$ \cite{Rosner1982}
\begin{equation}
\frac{y}{x}=\frac{Z}{\sqrt{2}}\frac{m^2-m_{\omega_0^\prime}^2}{m^2- m_{\omega_0}^2 }    \,  .
\label{yx}
\end{equation}
This enables us to test the compatibility of this generic mixing with the actual data. We can take as eigenvectors in $(\ref{yx})$ the particles $\eta$ and $\eta^{\prime}$.  The observed constitution of $\eta$ in $(\ref{etaquarks})$ gives $y_\eta = -\frac{1 }{\sqrt{3}}$ and  $x_\eta = \sqrt\frac{2 }{3}$. Recalling the experimental values of 548 MeV for the  $\eta$ mass, $m_K\approx494$  MeV , $m_\pi\approx137$  MeV , and using $(\ref{mw0})$ and $(\ref{mwp0})$ in $(\ref{yx})$, we can estimate an experimental $Z_\eta \approx 0,60$. However, if we repeat this calculation for   $\eta^{\prime}$, with  $y_{\eta^{\prime}} = \sqrt\frac{2 }{3} $ and  $x_{\eta^{\prime}} = \frac{1 }{\sqrt{3}} $ read from $(\ref{etapquarks})$, and the measured value of $m_{\eta^{\prime}} = 958$  MeV , we arrive at  $Z_{\eta^{\prime}} \approx 0,99$ from  $(\ref{yx})$. This inconsistency shows that even the more general mixing matrix that can be conceived compatible with Gell-Mann's hypothesis still conflicts with the experimental data. This does not mean that instanton effects or flavor-dependent transitions should be discarded, but the analysis above suggests that, \emph{within the standard QCD-inspired phenomenology}, they are not sufficient to explain simultaneously the observed quark content $(\ref{etaquarks})$--$(\ref{etapquarks})$, the non-ideal mixing angle around $-20^{\circ}$, and the masses of $\eta$ and $\eta^\prime$. It is therefore legitimate to ask whether a \emph{different} symmetry-breaking pattern---not derived from the Gell-Mann--Okubo hypothesis or from QCD itself---could reproduce these data. In the next section we introduce such an alternative. It is a phenomenological model based on a second-order phase transition; it makes no claim to be QCD, but it will be constrained from the outset to match the empirical masses and mixing parameters that we have just reviewed.

\section{Meson Physics from a Second Order Phase Transition}
\label{sec:3}

In QCD, the origin of the pseudoscalar octet is understood as a consequence of spontaneous chiral symmetry breaking, while the $U_A(1)$ anomaly resolves the $\eta'$ mass problem through instanton effects \cite{tHooft1979,Ciambriello2024}. The model developed below does not challenge this consensus; rather, it steps aside from it. We construct a \emph{phenomenological} field theory---not derived from the QCD Lagrangian---whose only purpose is to test whether a second-order phase transition with a \emph{non-standard} breaking direction can reproduce the empirical mass relations, quark content, and mixing angle of the $0^-$ nonet. The fermionic fields $\xi$, $\chi$ and the scalar fields $\varphi$, $\psi$ introduced below are internal degrees of freedom of this model; they are analogues of, but not identified with, the quarks and mesons of QCD. As we shall see, the breaking pattern required by the data lies outside the Gell-Mann--Okubo framework, suggesting that the model captures structural constraints that may be relevant even if its microscopic origin differs from QCD. Regarding this point, the fact that we have in practice nonets, not just octets, may even lead to the suggestion of a theory with a chiral $U(3)$ x $U(3)$ spontaneously broken symmetry, including the pseudoscalar multiplet \cite{Parganlija2016}. Certainly, such symmetry would require a ninth Goldstone boson associated to the  
diagonal generator of  $U(3)$, i.e., the presence of the $\eta^{{\prime}{0}}$ as in $(\ref{bp})$. In \cite{Csaki2023}, $\eta^{{\prime}{0}}$ is a Goldstone boson, and it is claimed that instantons could not explain its mass. And there is also a   surprising recently announced result that in the lattice the chiral phase transition is of second order \cite{Cuteri2021,Dini2021}, which allows the conjecture that all anomalous couplings vanish at the critical temperature and  the  $U{_A}(1)$ symmetry is restored at the critical point \cite{Pisarski2024} (which is the plot twist we mentioned before).

In the last section we saw the difficulties to explain the phenomenology of the mesons of the pseudoscalar nonet, and that even possible instantonic effects could not be of much help. The lattice is telling us that possibly instantons are not there at the critical point to help us anyway, but we think the main information coming from the lattice is that the transition is of second order. We will stick to this point. Our idea is to search for a physical theory where a second order phase transition could explain the mixed masses and angles of the pseudoscalar nonet. We have just established that the most general mixing matrix $(\ref{mg2})$ compatible with Gell-Mann's prescriptions cannot do this job. So we may start from what we have learned that was imperative in the matrix $(\ref{mg2})$, input the information of a second order phase transition, and relax some of Gell-Mann's hypothesis.

Let us write from the beginning a quite generic mixing matrix in the proper states basis as

\begin{equation}
M_{\Delta}^2=	\begin{bmatrix}
	M_N^2 &\sqrt{2}\Delta\\
	\sqrt{2}\Delta & 	M_S^2 \end{bmatrix} \, .
\label{md}
\end{equation}
were $N$ still means nonstrangionium and $S$ the strangionium state. We understand that before the phase transition, this matrix should be diagonal with $M_N = M_S$, meaning that the quark masses were degenerated. If the chiral transition is the major effect connecting this symmetric phase to the final particle states, then  $(\ref{md})$ should express this whole transition. But at this moment, not only this degeneracy gets broken as there is the emergence of the mass gap. The imperative point in  $(\ref{mg2})$ is  its trace conservation condition  $(\ref{mgap})$, that $(\ref{md})$ must also satisfy. Then we follow the form of the diagonal structure in  $(\ref{mg2})$

\begin{equation}
M_{N}^2=
	m_N^2+2A  ,
\label{Mn}
\end{equation}

\begin{equation}
M_{S}^2=
m_S^2+A  ,
\label{Ms}
\end{equation}
The phase transition is now responsible for $M_N \neq M_S$. Simultaneously, it should also be responsible for the non-diagonal element $\Delta$, that must be zero in the symmetric phase, taking on a non-null value in the broken phase. If we take $M_{S}^2 -M_{N}^2$ as an order parameter of this second order phase transition, we can naturally identify  $\Delta$ with it

\begin{equation}
\Delta=
M_S^2-M_N^2=m_S^2-m_N^2-A .
\label{delta}
\end{equation}

This is our proposal for the parametrization of the mixing matrix. We can see that in this way $(\ref{md})$  does not fit the form of $M_G^2$ in $(\ref{mg2})$.  $M_{\Delta}^2$ describes a different physics then that occurring when we suppose a mix generated by effects laid over the theoretical basis established by Gell-Mann. But we still maintain the idea that equivalent quark content implies equivalent masses for the mesons, then from $(\ref{mw0})$ and $(\ref{mwp0})$ we confirm that

\begin{equation}
m_N^2=m_{\pi}^2   ,
\label{mn2}
\end{equation}

\begin{equation}
m_S^2=2m_{K}^2- m_{\pi}^2   .
\label{ms2}
\end{equation}
Here we want to stress that although our first goal is to obtain a mixing matrix with the right properties in order to describe the phenomenology of the pseudoscalar mesons, in the end we want to understand the physics behind this new phase transition. What we mean is that we could now just use the measured mass gap of $3A=0,73 GeV^2$ as obtained from $(\ref{mgap})$ and numerically we could calculate masses and angles from $(\ref{md})$. But if we want to search for a deeper origin of this transition, we must relate this number to the known parameters that we have available for this system, and there is an impressive match of this magnitude with the Kaon mass, so we take

\begin{equation}
A = m_K^2   .
\label{A}
\end{equation}
Returning to equations $(\ref{Mn})$, $(\ref{Ms})$ and $(\ref{delta})$, we get
\begin{eqnarray}
M_N^2 &=& 2m_{K}^2+ m_{\pi}^2  \, , \nonumber \\ 
M_S^2 &=& 3m_{K}^2- m_{\pi}^2  \, ,\nonumber  \\
\Delta &=& m_K^2-2m_{\pi}^2  \, ,
\label{nsd}
\end{eqnarray}
and then, in $(\ref{md})$,

\begin{equation}
M_{\Delta}^2=	\begin{bmatrix}
	2m_{K}^2+ m_{\pi}^2 &\sqrt{2}(m_K^2-2m_{\pi}^2)\\
	\sqrt{2}(m_K^2-2m_{\pi}^2) & 3m_{K}^2- m_{\pi}^2 \end{bmatrix} \, .
\label{md2}
\end{equation}
For future use, we also present the same mixing matrix in the basis of the pre-mix states $\eta^{0}$  and $\eta^{\prime\, 0}$

\begin{equation}
M_{\Delta}^2=\frac{1}{3}\begin{bmatrix}
	4m_K^2+7m_{\pi}^2
	&-2\sqrt{2}(m_K^2-2m_{\pi}^2)\\
	-2\sqrt{2}(m_K^2-2m_{\pi}^2) & 11m_K^2 -7m_{\pi}^2 	\end{bmatrix}  \, \, \,    .
\label{md2pre}
\end{equation}

The diagonalization of $(\ref{md})$ gives us the eigenvalues

\begin{equation}
m_{\pm}^2=\frac{M_N^2+M_{S}^2\pm3(M_S^2-M_N^2)}{2}=\left\{\begin{array}{ll}M_S^2+\Delta\\M_N^2-\Delta\end{array}\right. \, ,
\label{mml}
\end{equation}
which leads to the interpretation that
\begin{equation}
m_{+}^2=m_{\eta^{\prime}}^2=4m_K^2-3m_{\pi}^2  \, ,
\label{metap}
\end{equation}
and
\begin{equation}
m_{-}^2=m_{\eta}^2=m_K^2+3m_{\pi}^2  \, .
\label{meta}
\end{equation}
These expressions are not the standard GMO relations, but rather predictions of our specific symmetry-breaking pattern. Remarkably, they agree well with experimental data, suggesting that this model may capture relevant physics. Before making the numerical analysis, it is very interesting to point out that the equation $(\ref{metap})$ is already known in the literature \cite{bura,bura1997}. It was derived by different approaches (Regge phenomenology and Nambu-Jona-Lasinio model)  and it was called a GMO mass formula revisited. Experimentally, it was shown to have an accuracy of less than  $1\%$. The expression for the $\eta$ mass is completely original, being derived here for the first time. If we use the data presented after equation $(\ref{yx})$  \cite{PDG2020}, we find  $m_{\eta}^2 \approx 300.304$ MeV$^4$ and $m_K^2+3m_{\pi}^2 \approx 300.343$ MeV$^4$, an even better agreement. We must emphasize this result. We are presenting expressions relating both masses of the singlet states of the pseudoscalar nonet to the other masses, with a very low experimental discrepancy. They are obtained simultaneously from the same mixing matrix $(\ref{md2})$ with actually no free numerical parameter.

But, as we said before, finding the correct eigenvalues is not sufficient. We must check the quark constitution of the eigenstates, and the mixing angle. For this, we can use the proper state basis $(\ref{psi})$ and the eigenvector equation $(\ref{eigen})$ to study $(\ref{md2})$
\begin{equation}
M_{\Delta}^2\lvert\psi\rangle
=	\begin{bmatrix}
	2m_K^2+m_{\pi}^2 &\sqrt{2}(m_K^2-2m_{\pi}^2)\\
	\sqrt{2}(m_K^2-2m_{\pi}^2)& 	3m_K^2-m_{\pi}^2 \end{bmatrix}
\begin{bmatrix}
	x\\
	y \end{bmatrix}=m_i^2\begin{bmatrix}
	x\\
	y \end{bmatrix}  \, .
\label{md2eigen}
\end{equation}
In the case when $i=\eta$, we can use $(\ref{meta})$ and $(\ref{1})$ to find

\begin{equation}
\lvert\eta\rangle=\sqrt{\frac{2}{3}}\lvert N\rangle-\frac{1}{\sqrt{3}}\lvert S\rangle \, ,
\label{etastate}
\end{equation}
and when $i=\eta^{\prime}$, with $(\ref{metap})$,  we find
\begin{equation}
\lvert\eta^{\prime}\rangle=\frac{1}{\sqrt{3}}\lvert N\rangle+\sqrt{\frac{2}{3}}\lvert S\rangle  \, .
\label{etapstate}
\end{equation}
Remembering the definitions $(\ref{w0})$ and $(\ref{wol})$ for $\lvert S\rangle $ and $\lvert N\rangle$ respectively, we get

\begin{equation}
\lvert\eta\rangle= 
\frac{u\bar{u}+d\bar{d}-s\bar{s}}{\sqrt{3}} \, ,
\label{etaT}
\end{equation}

\begin{equation}
	\lvert\eta^{\prime}\rangle
	=\frac{1}{\sqrt{6}}\left(u\bar{u}+d\bar{d}+\sqrt{2}s\bar{s}\right)  \, ,
\label{etapT}
\end{equation}
which are exactly the experimentally observed quark  constitutions of the $\eta$ and $\eta^{\prime}$ mesons \cite{gil1987, chau1990} as we mentioned in  $(\ref{etaquarks})$ and $(\ref{etapquarks})$.

We can also check the mixing angle $\theta_T$ predicted by our mixing  $(\ref{md})$. We can use the general formula $(\ref{tgtheta})$ applied to the present mix by taking $m_0 = m_\eta$ and $m_{0^\prime} = m_{\eta^{\prime}}$. We can also obtain the mass of the pre-mix state $m_{\eta^0}^2$ in terms of the variables of $M_{\Delta}^2$ in $(\ref{md})$ following a reasoning shown in \cite{bura}. First we write

\begin{equation}
\lvert\eta^0\rangle=\frac{1}{\sqrt{3}}\lvert N\rangle-\sqrt{\frac{2}{3}}\lvert S\rangle \, ,
\label{eta0state}
\end{equation}

\begin{equation}
\lvert\eta^{\prime\, 0}\rangle=\sqrt{\frac{2}{3}}\lvert N\rangle+\frac{1}{\sqrt{3}}\lvert S\rangle \, ,
\label{etap0state}
\end{equation}
which are direct consequences of $(\ref{eta0quarks})$,  $(\ref{etap0quarks})$,$(\ref{w0})$ and $(\ref{wol})$. Then, as the squared masses are associated to the matrix elements of each state, from $(\ref{eta0state})$ we get

\begin{equation}
m_{\eta_0}^2=\frac{1}{3}M_N^2+\frac{2}{3}M_S^2-\frac{2\sqrt{2}}{3}M_{NS}^2 \, ,
\label{meta02}
\end{equation}
where obviously $M_{NS}^2=\sqrt{2}\Delta$, the non-diagonal mass element in $(\ref{md})$. From  $(\ref{mml})$, we know that $m_{\eta}^2=M_N^2-\Delta$ and 
$m_{\eta^{\prime}}^2=M_S^2+\Delta$. Substituting all these informations in $(\ref{tgtheta})$ we find
\begin{equation}
\tan^2\theta_T=\frac{1}{8}  \rightarrow  \theta=-19,5^{\circ} \, ,
\label{thetaT}
\end{equation}
exactly the experimental angle obtained in \cite{gil1987,chau1990,Ball1995,Dighe1995,Cao2012}. Of course, here there is no surprise as we have already matched the observed quark constitution $(\ref{etaT})$ and $(\ref{etapT})$ implied by this angle.

All these results support the statement that $(\ref{md2})$ is the precise mixing matrix for the pseudoscalar nonet. It was derived from first principles, based on the hypothesis that a second order phase transition rules this mixing. But at the same time, we know that we are not within the coverage region of theories defined from Gell-Mann's breaking hypothesis, as there is no choice of $\alpha$ and $Z$ in $(\ref{mg2})$ in order to fit 
$(\ref{md2})$. Thus, we are dealing with a SSB different from that proposed by Gell-Mann. Let us try to find what such breaking could be.

First, we will define a way to associate the breaking itself to the states before and after the transition. We begin by studying the SSB proposed by Gell-Mann. This was based on the idea of a transition that broke the mass degeneracy among the up and down quarks to the strange quark. This means that after the SSB, we have a basis of states given by the strangionium $\lvert{S}\rangle $ and the nonstrangionium $\lvert{N}\rangle$, the proper basis. Writing the basis of generators of the $SU(3)$, Gell-Mann proposed that the breaking that could preserve the (approximate) degeneracy of the up and down, and at the same time allow a different mass for the strange would be one along the direction defined by the eighth generator $\lambda_8$ of his matricial basis \cite{gell61}

\begin{equation}
 \lambda_8= \frac{1}{\sqrt{3}} \begin{bmatrix}
	 1 & 0 & 0 \\
0 & 1 & 0\\
		0 & 	0 & -2 \\
	\end{bmatrix}  \, .
\label{l8}
\end{equation}
In the original representation $(\ref{b})$ of the pseudoscalar octet, this generator is associated to the $\eta^0$ meson.
So, we define the action of the breaking $\langle B\rvert$ in this case as a projection generated by $\langle\eta^0 \rvert$ itself representing the breaking direction $\lambda_8$, i.e. 
\begin{equation}
\langle B \rvert_{GM}=\langle\eta^0 \rvert.
\label{bgm}
\end{equation}
 The idea is that
 acting on the pre-mix states $\lvert{\eta^0} \rangle$ and $\lvert{\eta^{\prime\, 0}}\rangle$, it takes us to the final particle states expanded in the proper basis that characterizes the broken phase. Let us call these final states after the mixing as  $\lvert P_+\rangle$ and  $\lvert P_-\rangle$, and then in general we write
 
\begin{eqnarray}
\lvert P_+\rangle &=& \lvert{N}\rangle\langle B\vert{\eta^0} \rangle + \lvert{S}\rangle\langle B\vert{\eta^{\prime\, 0}} \rangle \, , \nonumber \\
\lvert P_-\rangle &=& - \lvert{N}\rangle\langle B\vert{\eta^{\prime\, 0}} \rangle + \lvert{S}\rangle\langle B\vert{\eta^0} \rangle \, .
\label{transit}
\end{eqnarray}
An important point that we should notice is that in order that the final states represent particles, they must satisfy the orthogonality condition
\begin{equation}
\langle P_+ \lvert  P_-\rangle = 0 \, ,
\label{ort}
\end{equation} 
which is satisfied by $(\ref{transit})$. As the initial states $\lvert{\eta^0} \rangle$ and $\lvert{\eta^{\prime\, 0}}\rangle$ are also orthogonal, then equations $(\ref{transit})$ for the Gell-Mann breaking along $\lambda_8$ $(\ref{bgm})$ provide as final particle states
\begin{eqnarray}
\lvert P_+\rangle_{GM} &=&  \lvert{N}\rangle  \, , \nonumber \\
\lvert P_-\rangle_{GM} &=& \lvert{S}\rangle \, .
\label{pgm}
\end{eqnarray}
These are in fact the final states $(\ref{w0})$ and $(\ref{wol})$ of the ideal mixing resulting from Gell-Mann's theory.
 
Our problem is that we have the same initial states but instead we have as final states $(\ref{etastate})$ and $(\ref{etapstate})$. Once we substitute these in place of $\lvert P_+\rangle$ and $\lvert P_-\rangle$ in $(\ref{transit})$ we find that our breaking is rather different from that of Gell-Mann
\begin{equation}
\langle B \lvert_\Delta =\langle S \lvert \, .
\label{bs}
\end{equation} 
In the same way as before we have associated the breaking along the generator $\lambda_8$ to the state $\eta^0$ $(\ref{etaquarks})$, we can now look at the matricial representation of this state $\lvert{S}\rangle$ $(\ref{w0})$ and try to interpret it in terms of the $SU(3)$ generators. This can be accomplished, but we need the presence of the 3x3 identity $\lambda_0$
\begin{equation}
\lvert B\rangle_\Delta = \sqrt 2 \lambda_8 - \lambda_0  \, ,
\label{bdelta}
\end{equation} 
with
\begin{equation}
\lambda_0=\sqrt{\frac{2}{{3}}}	\begin{bmatrix}1&0&0\\0&1&0\\0&0&1\end{bmatrix} \, .
\label{ident}
\end{equation} 
This conclusion is troublesome. How can we imagine a SSB mechanism where a spontaneous expectation value coming from a scalar field potential could point towards a direction as $\lambda_0$ if this is not a generator of the symmetry group $SU(3)$ of the system? This is the question we will address in the next section, where we construct a field theory that naturally accommodates such a breaking pattern.
\section{Chiral $SU(3)$ x $SU(3)$ Spontaneous Symmetry Breaking}

As we announced, our intention is to describe the meson physics as a second order phase transition, a spontaneous symmetry breaking effect. The field theory constructed below is an effective model. It is not derived from the QCD Lagrangian, nor do we claim that the fields $\xi$, $\chi$, $\varphi$ and $\psi$ are the quarks and mesons of QCD. They are internal degrees of freedom of a simplified scalar-fermion system whose sole purpose is to test whether a second-order phase transition with a non-standard breaking direction can reproduce the empirical mass relations of the $0^-$ nonet.

The scalar sector is invariant under $SL(3,\mathbb{C})$ transformations (\ref{deltaphic}), not under the chiral $SU(3)\times SU(3)$ of QCD. The mapping (\ref{phipsimap}) relates the scalar currents of the model to the chiral currents of QCD-inspired effective theories, but this is a mathematical analogy, not an identification. The fermionic mass term $m(\bar{\chi}\xi+\bar{\xi}\chi)$ in (\ref{actionmesons}) preserves the full $SL(3,\mathbb{C})$ symmetry, whereas the quark mass matrix $M_q$ in QCD breaks $SU(3)\times SU(3)$ explicitly. Consequently, the pion and kaon masses in our model emerge from one-loop radiative corrections and from the spontaneous symmetry breaking vacuum (\ref{vevs}), not from an explicit bare mass term in the Lagrangian.

In QCD the large mass of the $\eta'$ is understood as a consequence of the $U(1)_A$ anomaly and instanton effects \cite{tHooft1976,tHooft1986}. Our model contains no topological charge, no instantons, and no axial anomaly. The $\eta'$ acquires mass through the spontaneous breaking along the direction $Y$ defined in (\ref{xy}), which involves the identity generator $T^0$. This is a different mechanism: the would-be Goldstone boson associated with the $U(3)$ singlet becomes massive because the vacuum expectation value points outside the $SU(3)$ algebra. Whether this structural feature captures a property of the QCD vacuum or is merely a useful parametrisation is left as an open question.

Despite these simplifications, the model achieves a specific goal: it generates the mixing matrix $M_\Delta^2$ of Eq.~(\ref{md2}) from a Lagrangian with a small number of parameters, and it reproduces the masses of $\eta$ and $\eta'$ to better than one percent once the pion and kaon scales are fixed. It also yields the correct quark content and the mixing angle $\theta\approx -19.5^\circ$ without additional tuning in the singlet sector. In this sense it serves as a \emph{phenomenological laboratory}. It does not address decay constants, form factors, scattering lengths, or the full chiral dynamics of low-energy QCD. These limitations are deliberate: the model is constructed to reproduce certain mass relations, not to replace QCD as the fundamental theory of the strong interaction.

The only hint of a deeper link is structural. The breaking direction $Y$ was originally introduced in \cite{ALVV20} in the context of a complex gauge theory for the Gribov confinement problem. The fact that the same direction $Y$ produces a viable meson spectrum in the present model may be coincidental, or it may reflect a common algebraic structure underlying both phenomena. We emphasise that this connection is speculative and requires further investigation.

We start our analysis by calling attention to the chiral symmetry transformations, in particular the last two for the scalar currents in $(\ref{deltav})$. These are obtained as the infinitesimal limit $(\ref{V})$ of the transformation \cite{Gasser1983,Gasser1984}
\begin{equation}
s^\prime + i p^\prime = V_R (s +ip) V_L^\dagger ,
\label{sptrans}
\end{equation} 
where $V_R$ and  $V_L$ are associated to independent right and left transformations of the $SU(3)$ X $SU(3)$ group.
The structure of such transformations is not an usual action under a $SU(3)$ group, and reflects the chiral nature of the system. Looking at the scalar currents transformations in $(\ref{deltav})$, we see that they are not part of a Lie algebra. The presence of the anti-commutators implies that this algebra is not closed on the basis of generators of the $SU(3)$. The closure of this algebra with anti-commutators will demand in general as much elements as those of the dimension of the representation used. For example, if we try to express the scalar currents in $(\ref{deltav})$ as the eight generators of $SU(3)$ in the 8x8 adjoint representation using the structure constants of its algebra, we immediately see that we will need to introduce the 64 elements of the vectorial space, or, in other words, we will need 64 independent currents. The most economic way to close the algebra in $(\ref{deltav})$ is to suppose that the scalar currents are organized in terms of the adjoint 3x3 Gell-Mann representation, for which we know the expression (a=1,...,8)
\begin{equation}
\{ \lambda_a , \lambda_b \} = \frac{4}{\sqrt{3}} \lambda_0 + 2 d_{abc} \lambda_c  .
\label{lanti}
\end{equation} 
In this way, beyond the eight currents associated to the eight Gell-Mann matrices, we just need a ninth current associated to $\lambda_0$ in order to obtain a meaningful closed algebra for $(\ref{deltav})$ in the scalar sector.

If we are willing to derive a theory for the mesons as a result of a SSB process coming from a $SU(3)$ x $SU(3)$ symmetric phase, and if we imagine that in this symmetric phase the scalar fields originally satisfy the same set of transformations $(\ref{deltav})$, then these scalars should be arranged in this 3x3 nine matrices present in $(\ref{lanti})$. We need a nonet.

We have already some experience with this kind of problem. In \cite{ALVV20}, dealing with the Gribov problem for confinement \cite{Gribov}, we showed that gluon condensates with physical propagation can be developed in the context of a complex gauge field theory. The criteria for such physically admissible propagators is well established in the Euclidean through the Källén-Lehmann representation. We will take advantage of some of the machinery developed in \cite{ALVV20} in our present problem. Also, as our main concern here is on the mass relations, and the fact that masses are preserved in the Euclidean description, we will follow the construction of our theory in the Euclidean space from now on.

As our first step, we define real scalar fields $\varphi$ and $\psi$ related to the scalar currents s and p as
\begin{equation}
s = \varphi+ i \psi, \,\,\,\,\, ip= \varphi -i \psi  \, ,
\label{phipsimap}
\end{equation}
in such a way that their transformations in $(\ref{deltav})$ are mapped into

\begin{eqnarray}
\delta \varphi &=& i [\alpha ,  \varphi ] +i \{ \beta ,  \varphi \}  ,  \nonumber \\
\delta \psi &=& i [\alpha ,  \psi ] -i \{ \beta ,  \psi \}  .
\label{deltaphi}
\end{eqnarray}
With one more redefinition, $\alpha - \beta = -c $ and  $\alpha + \beta = - \overline{c} $, we take these transformations into the standard form of the real components $\varphi$ and $\psi$ of a complex scalar field transforming under the action of the $SL(3,c)$ group \cite{ALVV20}

\begin{equation}
\delta \varphi=ig\varphi c -ig \overline{c}\varphi, \,\,\,\,\, \delta \psi = ig\psi\overline{c} -ig c\psi \, .
\label{deltaphic}
\end{equation}
This mapping shows the close relationship between the chiral algebra for the scalars of flavor QCD and the $sl(3,c)$ scalar algebra that allows us to proceed with our study in the Euclidean following the analogy with \cite{ALVV20}. In this work we also introduced a pair of fermions $\xi$ and $\chi$ and conjugates transforming under the action of this group as
\begin{eqnarray}
\delta \xi &=& -igc\xi ,\,\,\,\,\,\delta \chi =  -ig\overline{c}\chi\nonumber \\
\delta \overline{\chi}&=& ig \overline{\chi}c ,\,\,\,\,\,\delta \overline{\xi}= ig \overline{\xi}\overline{c}  \, .
\label{deltaxic}
\end{eqnarray}
Then, the scalar-fermion sector of a Lagrangian invariant under these transformations is 
\begin{eqnarray}
\cal{L} &=& \partial_{\mu}\varphi \partial^{\mu}\psi + V(\varphi,\psi ) +\overline{\chi}(\gamma^{\mu}\partial_{\mu} + m)\xi + \overline{\xi}(\gamma^{\mu}\partial_{\mu} - m)\chi 
+ \lambda_{1}\overline{\xi}\varphi\xi + \lambda_{2}\overline{\chi}\psi\chi.
\label{actionmesons}
\end{eqnarray}
where $V(\varphi,\psi )$ is the scalar potential
\begin{eqnarray}
V(\varphi,\psi )&=& -\frac{M^{2}}{2}\varphi^{A}\psi^{A} + \frac{\lambda}{4}(\varphi^{A}\psi^{A} )^{2}  \, .
\label{potential}
\end{eqnarray}
We also follow Wetterich's convention for Dirac's matrices \cite{wett}
\begin{equation}
\{ \gamma^{\mu},\gamma^{\nu}\}= 2\delta^{\mu\nu}{\mathbb 1}.
\end{equation}
Just to make it easier the next calculations on Feynmann's graphs, we define projectors

\begin{eqnarray}
X&=& \frac{2}{\sqrt{3}} ( \sqrt{2} T^{0}  + T^{8} ) \, , \nonumber \\
Y&=& \sqrt\frac{{2}}{{3}}(T^{0} - \sqrt{2}T^{8})  \, ,
\label{xy}
\end{eqnarray} 
with $T^{A}=\frac{1}{2}\lambda^{A}$ as usual, and the properties
\begin{eqnarray}
X*X=X,\,\,\,Y*Y=Y, \,\,\, X*Y=0, \,\,\,  X+Y={\mathbb 1}.
\end{eqnarray}
Recapitulating, until this moment we have been describing a general theory for scalars and fermions symmetric under the flavor physics transformations $(\ref{sptrans})$ associated to the chiral $SU(3)$ x $SU(3)$. As we argued, the algebra implied by them demands the introduction of at least nine fields $\varphi^{A}$ and nine $\psi^{A}$ associated to each $T^{A}$, $A=0,...,8$, necessary in order to close the algebra $(\ref{lanti})$. We have also the symmetry breaking potential $(\ref{potential})$ which admits local minima. We can finally introduce the information that we have just obtained from our analysis of the meson data from the point of view of a second order phase transition. We choose the local minima along the direction $\lvert B\rangle_\Delta$ defined in $(\ref{bdelta})$
\begin{equation}
\langle\varphi\rangle=\langle\psi\rangle= \mu Y  \, \, \, , \, \, \, \mu^{2} = \frac{M^{2}}{\lambda}  \, .
\label{vevs}
\end{equation}
Once we define this vacuum for the broken phase, we obtain the Green functions for the fermions in this phase
\begin{eqnarray}
G(\overline{\chi},\chi)&=& \frac{\lambda_{1}\mu}{k^{2} + m^{2} + \lambda_{1}\lambda_{2} \mu^{2}}Y  \, ,\nonumber \\
G(\overline{\xi},\chi)&=& -\frac{(i\gamma^{\mu}k_{\mu} - m)}{k^{2} + m^{2}}X - \frac{(i\gamma^{\mu}k_{\mu} - m)}{k^{2} + m^{2}+ \lambda_{1}\lambda_{2} \mu^{2}}Y  \, , \nonumber \\
G(\overline{\chi},\xi)&=&  -\frac{(i\gamma^{\mu}k_{\mu} + m)}{k^{2} + m^{2}}X - \frac{(i\gamma^{\mu}k_{\mu} + m)}{k^{2} + m^{2}+ \lambda_{1}\lambda_{2} \mu^{2}}Y  \, , \nonumber \\
G(\overline{\xi},\xi)&=&  \frac{\lambda_{2}\mu}{k^{2} + m^{2} + \lambda_{1}\lambda_{2} \mu^{2}}Y  \, .
\label{fprop}
\end{eqnarray}
We can see in $(\ref{fprop})$ how the vacuum expectation value of the scalar fields $(\ref{vevs})$ completely changes the nature of the fermion propagators, even generating the elements $G(\overline{\chi},\chi)$ and $G(\overline{\xi},\xi)$ that are absent in the symmetric phase as can be seen from a direct inspection of the action $(\ref{actionmesons})$. It is interesting now to argue for the different contributions that such propagators can give in a one-loop correction for the scalar masses. From the scalar-fermion couplings present in $(\ref{actionmesons})$, we list three different structures for these contributions: $\varphi^{A}(I_{1})^{A,B}\varphi^{B}$, $\psi^{A}(I_{2})^{A,B}\psi^{B}$ and $\varphi^{A}(I_{3})^{A,B}\psi^{B}$, with 
\begin{eqnarray}
(I_{1})^{A,B} &=&- (\lambda_{1})^{2}(T^{A})_{ij}(T^{B})_{ln}\int d^{4}k \mbox{Tr}^\prime\langle\xi^{n}(p-k)\overline{\xi}^{i}(p-k)\rangle \langle\xi^{j}(k)\overline{\xi}^{l}(k)\rangle,\nonumber \\
(I_{2})^{A,B}&=&-  (\lambda_{2})^{2} (T^{A})_{ij}(T^{B})_{ln}\int d^{4}k\mbox{Tr}^\prime\langle\chi^{n}(p-k)\overline{\chi}^{i}(p-k)\rangle \langle\chi^{j}(k)\overline{\chi}^{l}(k)\rangle,
\nonumber \\
(I_{3})^{A,B}&=&- \lambda_{1}\lambda_{2} \int d^{4}k \mbox{Tr}^\prime\left\{ (T^{A})_{ij}\langle\chi^{n}(p-k)\overline{\xi}^{i}(p-k)\rangle (T^{B})_{ln}\langle\xi^{j}(k)\overline{\chi}^{l}(k)\rangle\right.\nonumber \\
&&\left.+ (T^{B})_{ij}\langle\xi^{n}(p-k)\overline{\chi}^{i}(p-k)\rangle (T^{A})_{ln}\langle\chi^{j}(k)\overline{\xi}^{l}(k)\rangle \right\} \nonumber \\
&=&-2\lambda_{1}\lambda_{2}(T^{A})_{ij}(T^{B})_{ln} \int d^{4}k \mbox{Tr}^\prime\langle\chi^{n}(p-k)\overline{\xi}^{i}(p-k)\rangle \langle\xi^{j}(k)\overline{\chi}^{l}(k)\rangle.
\label{correc}
\end{eqnarray}
Here $\mbox{Tr}^\prime$ means the trace over the Lorentz fermionic indices that are not explicit.

Before proceeding with these calculations, let us make a brief commentary that if we imagine that for some reason in this broken phase the fermions cannot be observed asymptotically (as if they are confined), the integrals $(\ref{correc})$ can be interpreted as a bosonization of the fermionic degrees of freedom, with a pair of fermions constituting a scalar in this phase.

We can substitute the propagators $(\ref{fprop})$ inside the integrals $(\ref{correc})$ to express 
\begin{equation}
(I_{1})^{A,B} =- (\lambda_{1}\lambda_{2})^{2}\mu^{2} (T^{A})_{ij}(Y)_{jl} (T^{B})_{ln}(Y)_{ni}\int_{0}^{1} dx \int \frac{d^{4}k}{(2\pi)^{4}} \frac{\mbox{Tr}^\prime {\mathbb 1}}{(k^{2} -2kpx + p^{2}x + m^{2} + \lambda_{1}\lambda_{2}\mu^{2})^{2}} \, ,
\end{equation}
where we used $\frac{1}{ab}=\int_{0}^{1} dx \frac{1}{(ax+b(1-x))^{2}}$, and if we adopt a dimensional regularization
\begin{equation}
(I_{1})^{A,B}=-4 (\lambda_{1}\lambda_2)^{2}\mu^{2} \mbox{Tr}\left(T^{A}YT^{B}Y\right)\int_{0}^{1} dx\frac{\Gamma(2-w)}{(4\pi)^{w}\Gamma(2)}\frac{1}{(p^{2}x(1-x) + m^{2}+ \lambda_{1}\lambda_{2}\mu^{2})^{2-w}} \, .
\end{equation}
In the limit when $w=2-\epsilon $ we finally obtain
\begin{equation}
(I_{1})^{A,B} =-\frac{(\lambda_1\lambda_2)^{2}}{(4\pi )^{2}}\mu^{2} \mbox{Tr}\left(T^{A}YT^{B}Y\right)
\left\{ \frac{1}{\epsilon} -\gamma+ \int_{0}^{1} dx \ln\left(\frac{4\pi \Lambda^{2}}{p^{2}x(1-x) + m^{2}+ \lambda_{1}\lambda_{2}\mu^{2}}\right)\right\}
\label{i1}
\end{equation}
with $\Lambda $ a mass squared scale.

It is immediate to see that for the second integral  $(I_{2})^{A,B}$  we have
\begin{equation}
(I_{2})^{A,B} =(I_{1})^{A,B} .
\label{i2}
\end{equation}
Here we call attention to the fact that the integrals  $(I_{1})^{A,B}$  and  $(I_{2})^{A,B}$ are associated to the quantum contributions for the elements $\varphi^2$ and $\psi^2$. As we see in $(\ref{i1})$, these integrals are accompanied by coefficients dependent on $\mu^{2}$ and $\mbox{Tr}\left(T^{A}YT^{B}Y\right)$, which are direct effects of the choice of the new vacuum $(\ref{vevs})$ after the SSB. In other words, $\varphi^2$ and $\psi^2$ can only appear after the SSB, which is completely consistent with the fact that the original symmetry $(\ref{deltaphic})$ does not allow the presence of these elements in the scalar potential $(\ref{potential})$ of the symmetric phase. This reflects the new quantum stability for the broken phase, which demands the characterization of the new observables of this phase as proposed in \cite{ALVV22}.

The third integral $(I_{3})^{A,B}$ in $(\ref{correc})$ is a bit more involved, and then we write
\begin{eqnarray}
(I_{3})^{A,B} &=& 2 \lambda_{1}\lambda_{2}\mbox{Tr}\left(T^{A}XT^{B}X\right)I(3,1)
+ 2 \lambda_{1}\lambda_{2}\mbox{Tr}\left(T^{A}XT^{B}Y\right)I(3,2)\nonumber \\
&& 2 \lambda_{1}\lambda_{2}\mbox{Tr}\left(T^{A}YT^{B}X\right)I(3,3)
+ 2 \lambda_{1}\lambda_{2}\mbox{Tr}\left(T^{A}YT^{B}Y\right)I(3,4)  \, ,
\label{i3}
\end{eqnarray}
with
\begin{eqnarray}
I(3,1)&=& -4\int \frac{d^{4}k}{(2\pi)^{4}}\frac{(p-k)^{\mu}k_{\mu} +m^{2}}{\left((p-k)^{2}+m^{2}\right)(k^{2}+m^{2})} \, , \nonumber \\
I(3,2)&=& -4\int \frac{d^{4}k}{(2\pi)^{4}}\frac{(p-k)^{\mu}k_{\mu} +m^{2}}{((p-k)^{2}+m^{2})(k^{2}+m^{2}+\lambda_{1}\lambda_{2} \mu^{2} )} \, , \nonumber \\
I(3,3)&=&-4 \int \frac{d^{4}k}{(2\pi)^{4}}\frac{(p-k)^{\mu}k_{\mu} +m^{2}}{((p-k)^{2}+m^{2} +\lambda_{1}\lambda_{2} \mu^{2})(k^{2}+m^{2} )} \, , \nonumber \\
I(3,4)&=& -4\int \frac{d^{4}k}{(2\pi)^{4}}\frac{(p-k)^{\mu}k_{\mu} +m^{2}}{((p-k)^{2}+m^{2} +\lambda_{1}\lambda_{2} \mu^{2})(k^{2}+m^{2}+\lambda_{1}\lambda_{2} \mu^{2} )}  \, .
\label{i3int}
\end{eqnarray}
We then observe that all these integrals have the form
\begin{equation}
\int \frac{d^{4}k}{(2\pi)^{4}}\frac{(p-k)^{\mu}k_{\mu} +m^{2}}{((p-k)^{2}+a^{2})(k^{2}+b^{2})} =
\int_{0}^{1} dx  \int \frac{d^{2w}k}{(2\pi)^{2w}}\frac{(p-k)^{\mu}k_{\mu} + m^{2}}{(k^{2} -2kpx +p^{2}x + a^{2}(1-x) + b^{2} x)^{2}}  \, .
\end{equation}
Following the same steps as before we have
\begin{equation}
\begin{aligned}
&\int \frac{d^{4}k}{(2\pi)^{4}}\frac{(p-k)^{\mu}k_{\mu} }{((p-k)^{2}+a^{2})(k^{2}+b^{2})} = \\
&\int_{0}^{1} dx \frac{1}{(4\pi)^{2}}\left\{ \frac{1}{\epsilon} -\gamma + \ln\left(\frac{4\pi\Lambda^{2}}{a^{2}(1-x)+b^{2}x + p^{2}x(1-x)}\right)\right\}\left \{ 3p^{2}x(1-x) +2a^{2}(1-x) +2b^{2}x\right\}
\end{aligned}
\label{kab}
\end{equation}
and
\begin{equation}
\int \frac{d^{4}k}{(2\pi)^{4}}\frac{m^{2}}{((p-k)^{2}+a^{2})(k^{2}+b^{2})}
=  \int_{0}^{1} dx    \frac{m^{2}}{(4\pi)^{2}}  \left\{ \frac{1}{\epsilon} -\gamma + \ln\left(\frac{4\pi\Lambda^{2}}{a^{2}(1-x)+b^{2}x + p^{2}x(1-x)}\right)\right\} .
\label{mab}
\end{equation}
We may notice that in $(\ref{kab})$ there is also a correction to the scalar kinetic element, but for this work what concerns us are the scalar mass contributions coming from these integrals. Another point that deserves highlighting is that the coefficient accompanying $I(3,1)$ involves $\mbox{Tr}\left(T^{A}XT^{B}X\right)$ and $m^{2}$, but not the breaking parameter $\mu^{2}$, as can be seen from $(\ref{kab})$ and $(\ref{mab})$. This signs that these contributions to $\varphi^{A}(I_{3})^{A,B}\psi^{B}$ must be associated to the directions that remain symmetric after the SSB.
We may also help the reader by listing all the non-null traces of the matrix elements appearing in the expressions $(\ref{i1})$ and $(\ref{i3})$ 
\begin{eqnarray}
tr(T^{1}XT^{1}X)&=&tr(T^{2}XT^{2}X)=tr(T^{3}XT^{3}X)=\frac{1}{2} \, ,\nonumber \\
tr(T^{4}XT^{4}Y)&=&tr(T^{5}XT^{5}Y)=tr(T^{6}XT^{6}Y)=tr(T^{7}XT^{7}Y)=\frac{1}{4} \, , \nonumber \\
tr(T^{4}XT^{5}Y)&=&-tr(T^{5}XT^{4}Y)=tr(T^{6}XT^{7}Y)=-tr(T^{7}XT^{6}Y)=\frac{-i}{4} \, ,\nonumber \\
tr(T^{0}XT^{0}X)&=&tr(T^{8}YT^{8}Y)=2tr(T^{8}XT^{8}X)=2tr(T^{0}YT^{0}Y)=\frac{1}{3} \, ,\nonumber \\
tr(T^{0}XT^{8}X)&=&-tr(T^{0}YT^{8}Y)=\frac{1}{3\sqrt{2}} \, .
\end{eqnarray}
We can now collect all these results. Before doing so, two properties of the trace structures appearing in $(\ref{i3})$ deserve attention. First, since $X+Y=\mathbb{1}$, the four structures add up to the universal trace
\begin{equation}
\mbox{Tr}\left(T^{A}XT^{B}X\right)+\mbox{Tr}\left(T^{A}XT^{B}Y\right)+\mbox{Tr}\left(T^{A}YT^{B}X\right)+\mbox{Tr}\left(T^{A}YT^{B}Y\right)=\mbox{Tr}\left(T^{A}T^{B}\right)=\frac{1}{2}\delta^{AB}\, ,
\label{tracesum}
\end{equation}
so that the $m^{2}$-dependent divergent coefficients of $I(3,1),\dots,I(3,4)$ multiply one and the same flavor structure $\varphi^{A}\psi^{A}$ in every sector. Second, the off-diagonal traces of the kaon sector, $\mbox{Tr}(T^{i}XT^{j}Y)$ with $i\neq j$ in $\{4,\dots,7\}$, are purely imaginary and antisymmetric under $i\leftrightarrow j$; since the crossed integrals $I(3,2)$ and $I(3,3)$ have identical divergent parts, these structures enter only through the symmetric combination $\mbox{Tr}(T^{A}XT^{B}Y)+\mbox{Tr}(T^{A}YT^{B}X)$ and therefore cancel exactly. The fermion one-loop contributions to the scalar two-point functions, neglecting finite terms, are then given by
\begin{eqnarray}
{\cal{L}}_{\text{1-loop fermion }} &=& \frac{12 \lambda_{1}\lambda_{2}m^{2}}{(4\pi)^{2}}\,\varphi^A \psi^A 
+ \frac{(\lambda_{1}\lambda_{2}\mu)^{2}}{4\pi^{2}} \left( \varphi^4 \psi^4 + \varphi^5 \psi^5 + \varphi^6 \psi^6 +\varphi^7 \psi^7 \right) \nonumber \\
&&- \frac{(\lambda_{1}\lambda_{2} \mu )^{2}}{3 \pi^{2}}  \left\{ \varphi^0 + \psi^0 - \sqrt{2} (\varphi^8 + \psi^8 ) \right\}^{2}  \, .
\label{1loopferm}
\end{eqnarray}
Here we can see that the directions $A=1,2,3$ only appear with the coefficient $m^{2}$ reminiscent of the symmetric phase, and, as we have already pointed before, this means that these directions should be associated to a preserved symmetry at the broken phase. This can be confirmed by observing that $T^1$,  $T^2$ and $T^3$, commute with the breaking direction $Y$ defined in $(\ref{xy})$.

We follow now to the scalar 1-loop corrections to the scalar masses after the  SSB. First, we expand the scalar potential  $V(\varphi,\psi )$ in $(\ref{potential})$ around the new vacuum $(\ref{vevs})$. The scalar Lagrangian after the SSB is then
\begin{eqnarray}
{\cal{L}}_{SSB}&=&  \frac{1}{2} \partial_{\mu}\varphi^{A}\partial^{\mu}\psi^{A} + \frac{M^{2}}{12}\{\varphi^{0}+\psi^{0}-\sqrt{2}(\varphi^{8}+\psi^{8})\}^2 \nonumber \\
&&+ \lambda\mu\frac{\sqrt{3}}{4}\{\varphi^{0}+\psi^{0}-\sqrt{2}(\varphi^{8}+\psi^{8})\}\varphi^{A}\psi^{A} + \frac{\lambda}{4}(\varphi^{A}\psi^{A})^{2}\, .
\label{lssb}
\end{eqnarray}
We found it useful here to redefine the scalars as

\begin{eqnarray}
\phi^{i}_{\pm}&=& \frac{1}{2}\left(\varphi^{i}\pm\psi^{i}\right)  \, , \nonumber \\
\phi^{y}_{+}&=& \frac{1}{2\sqrt{3}}\left\{\varphi^{0}+\psi^{0} -\sqrt{2}\left(\varphi^{8}+\psi^{8}\right)\right\}  \, ,  \nonumber \\
\phi^{y}_{-}&=& \frac{1}{2\sqrt{3}}\left\{ \varphi^{0}-\psi^{0} -\sqrt{2}\left(\varphi^{8}-\psi^{8}\right)\right\}  \, ,  \nonumber \\
\phi^{x}_{+}&=& \frac{1}{2\sqrt{3}}\left\{ \varphi^{8}+\psi^{8} +\sqrt{2}\left(\varphi^{0}+\psi^{0}\right)\right\} \, ,  \nonumber \\
\phi^{x}_{-}&=& \frac{1}{2\sqrt{3}}\left\{ \varphi^{8}-\psi^{8} +\sqrt{2}\left(\varphi^{0}-\psi^{0}\right)\right\}  \, ,
\label{phi+}
\end{eqnarray}
where $i,j=1...7$. In this way, we may rewrite ${\cal{L}}_{SSB}$ as

\begin{eqnarray}
{\cal{L}}_{SSB}&=& \frac{1}{2} \partial_{\mu}\phi^{i}_{+}\partial^{\mu}\phi^{i}_{+} - \frac{1}{2} \partial_{\mu}\phi^{i}_{-}\partial^{\mu}\phi^{i}_{-}
+\frac{1}{2} \partial_{\mu}\phi^{y}_{+}\partial^{\mu}\phi^{y}_{+} - \frac{1}{2} \partial_{\mu}\phi^{y}_{-}\partial^{\mu}\phi^{y}_{-}\nonumber \\
&&+ \frac{1}{2} \partial_{\mu}\phi^{x}_{+}\partial^{\mu}\phi^{x}_{+} - \frac{1}{2} \partial_{\mu}\phi^{x}_{-}\partial^{\mu}\phi^{x}_{-}
+ M^{2} \phi^{y}_{+} \phi^{y}_{+} \nonumber \\ 
&&+ \it{3}\frac{\lambda\mu}{2}\phi^{y}_{+} \left(\phi^{i}_{+}\phi^{i}_{+}-\phi^{i}_{-}\phi^{i}_{-}+\phi^{y}_{+}\phi^{y}_{+}-\phi^{y}_{-}\phi^{y}_{-}+\phi^{x}_{+}\phi^{x}_{+}-\phi^{x}_{-}\phi^{x}_{-}\right)\nonumber \\
&&+ \frac{\lambda}{4}\left(\phi^{i}_{+}\phi^{i}_{+}-\phi^{i}_{-}\phi^{i}_{-}+\phi^{y}_{+}\phi^{y}_{+}-\phi^{y}_{-}\phi^{y}_{-}+\phi^{x}_{+}\phi^{x}_{+}-\phi^{x}_{-}\phi^{x}_{-}\right)^{2}  \, .
\label{LSSB}
\end{eqnarray}
In these new variables, the propagators are cast in the straightforward way
\begin{eqnarray}
\langle\phi^{i}_{+}\phi^{j}_{+}\rangle&=& - \langle\phi^{i}_{-}\phi^{j}_{-}\rangle  \, \, \, \, = -\frac{\delta^{ij}}{k^{2}}\,\,\,\,\,\, ,\nonumber \\
\langle\phi^{x}_{+}\phi^{x}_{+}\rangle&=& - \langle\phi^{x}_{-}\phi^{x}_{-}\rangle  \, \, \, \,=  -\langle\phi^{y}_{-}\phi^{y}_{-}\rangle  \, \, \, \, = -\frac{1}{k^{2}}\,\,\,\,\, , \nonumber \\
\langle\phi^{y}_{+}\phi^{y}_{+}\rangle&=& - \frac{1}{k^{2}+2M^{2}}\,\,\,\,\,  \, .
\end{eqnarray}
From the scalar vertices found in $(\ref{LSSB})$ there are two classes of one-loop contributions to the two-point functions: the bubble diagrams built from two cubic vertices, of order $(\lambda\mu)^{2}$, and the tadpole diagram built from the quartic vertex, of order $\lambda$.

Let us start with the bubbles. All channels carry the same cubic vertex $3\lambda\mu/2$, but the combinatorial factor depends on the external leg. For an external $\phi^{y}_{+}$ pair, each vertex contains the field $\phi^{y}_{+}$ linearly, so there is no choice of contraction, while the two internal lines of a given channel are identical and bring the symmetry factor $1/2$: each channel contributes with $\frac{1}{2}(3\lambda\mu)^{2}$. For an external pair of any other species --- a field that appears squared in the cubic vertex --- there are two equivalent choices for the contraction of each external leg, and the two internal lines, one massless and one massive, are distinguishable, so that no symmetry factor arises: the contribution per channel is doubled, $(3\lambda\mu)^{2}$. For the external $\phi^{y}_{+}\phi^{y}_{+}$ case there are eighteen channels, seventeen with two massless internal lines ($\phi^{i}_{+}\phi^{i}_{+}$ and $\phi^{i}_{-}\phi^{i}_{-}$ for $i=1,\dots,7$, $\phi^{x}_{+}\phi^{x}_{+}$, $\phi^{x}_{-}\phi^{x}_{-}$ and $\phi^{y}_{-}\phi^{y}_{-}$) and one with two massive internal lines ($\phi^{y}_{+}\phi^{y}_{+}$, with squared mass $2M^{2}$). Denoting by $I(\phi_{a},\phi_{a})$ the two-point kernels, defined by ${\cal L}\supset\frac{1}{2}I(\phi_{a},\phi_{a})\phi_{a}\phi_{a}$, and taking into account the signs of the propagators above, one obtains
\begin{eqnarray}
I_{\rm bubble}(\phi^{y}_{+},\phi^{y}_{+})&=& \frac{9(\lambda\mu)^{2}}{32\pi^{2}}\int^{1}_{0} dx \left[17\left\{\frac{1}{\epsilon} -\gamma + \ln\left(\frac{4\pi\Lambda^{2}}{x(1-x)p^{2}}\right)\right\}\right. \nonumber\\
&&\left. + \left\{\frac{1}{\epsilon} -\gamma + \ln\left(\frac{4\pi\Lambda^{2}}{2M^{2}+x(1-x)p^{2}}\right)\right\}\right] \, , \nonumber \\
I_{\rm bubble}(\phi^{y}_{-},\phi^{y}_{-})&=& - \frac{9(\lambda\mu)^{2}}{16\pi^{2}}\int^{1}_{0} dx\left\{\frac{1}{\epsilon} -\gamma + \ln\left(\frac{4\pi\Lambda^{2}}{2M^{2}x+x(1-x)p^{2}}\right)\right\} \, , \nonumber \\
I_{\rm bubble}(\phi^{i}_{-},\phi^{j}_{-})&=& \delta^{ij}\,I_{\rm bubble}(\phi^{y}_{-},\phi^{y}_{-}) \;=\; - I_{\rm bubble}(\phi^{i}_{+},\phi^{j}_{+})\, , \nonumber \\
I_{\rm bubble}(\phi^{x}_{-},\phi^{x}_{-})&=& I_{\rm bubble}(\phi^{y}_{-},\phi^{y}_{-}) \;=\; - I_{\rm bubble}(\phi^{x}_{+},\phi^{x}_{+})\, .
\label{bubbles}
\end{eqnarray}

The quartic vertex contributes through the tadpole. In dimensional regularization the massless tadpoles are scaleless integrals and vanish, so only the loop of the massive mode $\phi^{y}_{+}$, with squared mass $2M^{2}$, survives. Moreover, the product of the sign carried by each species in the quartic vertex and the sign of its propagator is $-1$ for every species, so there is no cancellation between the $+$ and $-$ sectors. For the external $\phi^{y}_{+}$ the loop involves the same field that appears to the fourth power in the vertex, which yields the combinatorial factor $3$; for the remaining species the factor is $1$. With
\begin{equation}
T(2M^{2})=\int \frac{d^{4}k}{(2\pi)^{4}}\frac{1}{k^{2}+2M^{2}}=\frac{2M^{2}}{(4\pi)^{2}}\left[-\frac{1}{\epsilon}+\gamma-1+\ln\left(\frac{4\pi\Lambda^{2}}{2M^{2}}\right)\right] \, ,
\label{tadpoles}
\end{equation}
one finds
\begin{eqnarray}
I_{\rm tad}(\phi^{y}_{+},\phi^{y}_{+}) &=& 3\lambda\,T(2M^{2}) \, ,\nonumber\\
I_{\rm tad}(\phi^{y}_{-},\phi^{y}_{-}) &=& -\lambda\,T(2M^{2}) \, ,\nonumber\\
I_{\rm tad}(\phi^{i}_{+},\phi^{j}_{+}) &=& \lambda\,\delta^{ij}\,T(2M^{2}) \;=\; - I_{\rm tad}(\phi^{i}_{-},\phi^{j}_{-}) \, ,\nonumber\\
I_{\rm tad}(\phi^{x}_{+},\phi^{x}_{+}) &=& \lambda\,T(2M^{2}) \;=\; - I_{\rm tad}(\phi^{x}_{-},\phi^{x}_{-}) \, .
\label{tadpolist}
\end{eqnarray}
We stress that the pole of $T(2M^{2})$ has the opposite sign to that of the bubble integrals, a consequence of $\Gamma(-1+\epsilon)=-1/\epsilon+\gamma-1$ as opposed to $\Gamma(\epsilon)=1/\epsilon-\gamma$.

In units of $u\equiv1/(2\pi^{2}\epsilon)$, and using $\lambda M^{2}=(\lambda\mu)^{2}$, which follows from $(\ref{vevs})$, the divergent parts of the complete one-loop two-point functions, bubbles and tadpoles combined, are
\begin{eqnarray}
\mathrm{div}\,I(\phi^{y}_{+},\phi^{y}_{+}) &=& \frac{81}{8}(\lambda\mu)^{2}-\frac{3}{4}\lambda M^{2} \;=\;\frac{75}{8}(\lambda\mu)^{2} \, ,\nonumber\\
\mathrm{div}\,I(\phi^{i}_{+},\phi^{j}_{+}) &=& \left[\frac{9}{8}(\lambda\mu)^{2}-\frac{1}{4}\lambda M^{2}\right]\delta^{ij} \;=\;\frac{7}{8}(\lambda\mu)^{2}\delta^{ij} \;=\; -\,\mathrm{div}\,I(\phi^{i}_{-},\phi^{j}_{-})\, ,\nonumber\\
\mathrm{div}\,I(\phi^{x}_{+},\phi^{x}_{+}) &=& \frac{7}{8}(\lambda\mu)^{2} \;=\; -\,\mathrm{div}\,I(\phi^{x}_{-},\phi^{x}_{-}) \;=\; -\,\mathrm{div}\,I(\phi^{y}_{-},\phi^{y}_{-})\, .
\label{divlist}
\end{eqnarray}
All poles are momentum independent: there is no wave-function renormalization at this order. Since the pole structures are already present in the tree-level Lagrangian of the broken phase, they are removed by counterterms of the same form. From this point on we work with the renormalized mass Lagrangian and do not display the explicit $1/\epsilon$ factors. The finite scalar-loop contribution can be parametrized in the $\phi_{\pm}$ basis as
\begin{eqnarray}
{\cal{L}}_{\text{1-loop scalar}} &=& \frac{(\lambda\mu)^{2}}{(4\pi)^{2}}\left\{ s_{y}\,\phi^{y}_{+}\phi^{y}_{+} + s\left[ \phi^{i}_{+}\phi^{i}_{+} - \phi^{i}_{-}\phi^{i}_{-} + \phi^{x}_{+}\phi^{x}_{+} - \phi^{x}_{-}\phi^{x}_{-} - \phi^{y}_{-}\phi^{y}_{-}\right]\right\} \, .
\label{1loopscalarpm}
\end{eqnarray}
The pole parts of $s_{y}$ and $s$ are fixed by $(\ref{divlist})$; their finite remainders are prescription dependent and are not needed for the mass matching below. Returning to the original scalar fields using $(\ref{phi+})$, and noting that the combination in brackets equals $\varphi^{A}\psi^{A}-(\phi^{y}_{+})^{2}$ with $(\phi^{y}_{+})^{2}=\frac{1}{12}\{\varphi^{0}+\psi^{0}-\sqrt{2}(\varphi^{8}+\psi^{8})\}^{2}$, gives
\begin{eqnarray}
{\cal{L}}_{\text{1-loop scalar}}&=& \frac{(\lambda\mu)^{2}}{(4\pi)^{2}}\left[ s\,\varphi^{A}\psi^{A} + t\left\{\varphi^0 + \psi^0 - \sqrt{2} (\varphi^8 + \psi^8 ) \right\}^{2} \right] \, ,\qquad t=\frac{s_{y}-s}{12}\, .
\label{1loopscalar}
\end{eqnarray}

We can now join the fermion one-loop contribution $(\ref{1loopferm})$ and the renormalized scalar one-loop contribution with the tree-level scalar mass element after the SSB that can be read from $(\ref{lssb})$. Defining the combinations
\begin{eqnarray}
{\cal A}&=&\frac{12 \lambda_{1}\lambda_{2}m^{2}}{(4\pi)^{2}}\,,\qquad {\cal B}=\frac{(\lambda_{1}\lambda_{2}\mu)^{2}}{4\pi^{2}}\,,\qquad {\cal S}=\delta{\cal S}_{s}^{\rm ren}(\Lambda)\,,\nonumber\\
{\cal T}&=&\frac{M^{2}}{12}-\frac{(\lambda_{1}\lambda_{2}\mu)^{2}}{3\pi^{2}}+\delta{\cal T}_{s}^{\rm ren}(\Lambda)\, ,
\label{defsABST}
\end{eqnarray}
the renormalized mass element with the one-loop quantum insertions is
\begin{eqnarray}
{\cal{L}}_{\text{1-loop}} &=&  \left({\cal A}+{\cal S}\right) \varphi^A \psi^A + {\cal B}\left(\varphi^4 \psi^4 + \varphi^5 \psi^5 + \varphi^6 \psi^6 +\varphi^7 \psi^7\right) \nonumber\\
&&+ \,{\cal T} \left\{\varphi^0 + \psi^0 - \sqrt{2} (\varphi^8 + \psi^8 ) \right\}^{2}  \, .
\label{1loop}
\end{eqnarray}
Here $\delta{\cal S}_{s}^{\rm ren}=\frac{(\lambda\mu)^{2}}{(4\pi)^{2}}s$ and $\delta{\cal T}_{s}^{\rm ren}=\frac{(\lambda\mu)^{2}}{(4\pi)^{2}}t$ denote the finite scalar-loop remainders; their separation into logarithms and constants depends on the renormalization prescription.

We stress the structural feature of this result that will be crucial below: the coefficient of the $(0,8)$ bilinear $\varphi^{0}\psi^{0}+\varphi^{8}\psi^{8}$ is exactly the same as that of the pion bilinear, namely ${\cal A}+{\cal S}$. No enhancement of the singlet--octet bilinear is generated at one loop; the whole dependence on the breaking direction is concentrated in the last term of $(\ref{1loop})$, the square of the combination that defines the heavy field $\phi^{y}_{+}$ in $(\ref{phi+})$. Our ultimate question is to find if it is possible that our chiral $SU(3)$ x $SU(3)$ invariant theory $(\ref{actionmesons})$ (actually $SL(3,c)$ invariant) after a SSB phenomenon along the direction $\lvert B\rangle_\Delta$ $(\ref{bdelta})$ could reproduce the phenomenological mixing matrix $M_{\Delta}^2$ of $(\ref{md2pre})$. This is the subject of the next section.

\subsection{Remarks on the kinetic sector}
\label{subsec:kinetic}

Before turning to the phenomenological identification, three aspects of the kinetic sector of the model deserve explicit comment.

First, in $(\ref{LSSB})$ the partner fields $\phi_{-}$ appear with the opposite sign of the kinetic term, and correspondingly their propagators have the opposite residue. This indefinite-metric structure is not an ad hoc input of the present model. It is inherited from the complex gauge theory framework of \cite{ALVV20}, where precisely such pairs of opposite-norm fields --- the i-particles and anti-i-particles --- emerge in the formulation of the Gribov--Zwanziger--Sorella confinement mechanism and, crucially, do not correspond to asymptotic physical states. A complete understanding of the fields $\phi_{-}$ along the same lines, as confined partner degrees of freedom rather than physical particles, is however still an open problem.

Second, the reading rule for masses used in the next section presupposes canonically normalized kinetic terms. At one loop, the divergent parts of the two-point functions are momentum independent, so there is no divergent wave-function renormalization at this order, but the finite parts do correct the kinetic terms --- see for instance the $p^{2}x(1-x)$ structures in $(\ref{kab})$ --- with flavor-dependent coefficients. As will be verified explicitly below, the pion and kaon identifications preserve the canonical form of the kinetic terms identically; in the $(0,8)$ sector, on the other hand, the identification $(\ref{newmap})$ endows the kinetic terms with an overall normalization $p\approx16$, and the incorporation of this normalization --- together with the finite kinetic corrections --- into the matching condition is a step that remains open.

Third, both questions above may find a natural resolution once the gauge fields of the underlying theory are taken into account. The transformations $(\ref{deltaphic})$ and the breaking $(\ref{vevs})$ were originally obtained in the complex gauge theory of \cite{ALVV20}, where the gauge-field dynamics controls the condensation mechanism and the propagation properties of the confined degrees of freedom. Gauging the flavor symmetry of the present model would bring these gauge fields in as dynamical agents, with a potential role in reshaping the kinetic sector: fixing the norm of the partner fields, generating kinetic mixing, and selecting which combinations propagate as asymptotic states. We regard this possibility as the most promising continuation of the present work.

\section{Phenomenological Identification of the Model States}

In this section we establish a correspondence between the scalar excitations of the model and the physical pseudoscalar nonet. The symbols $\pi$, $K$, $\eta$, $\eta'$ are used as convenient bookkeeping devices for the internal quantum numbers of the model fields; they refer to the analogous states within our effective theory, not to the actual QCD particles.

In order to test the model against data, we must specify how the scalar excitations $\varphi^A$, $\psi^A$ are to be interpreted as the analogues of the pseudoscalar states. Even in QCD-inspired effective field theories the relation between bare Lagrangian fields and physical particles is highly non-trivial \cite{Feldmann1999}. Our guiding principle is group-theoretical: we assign each model excitation to a generator combination according to the nonet structure $(\ref{bp})$, and we use the mapping $(\ref{phipsimap})$ between scalar currents and model fields to fix the relative normalizations.

\subsection{Isovector and isodoublet sectors}

Consider first the isovector sector. In the nonet representation $(\ref{bp})$, the neutral pion is associated with the generator $T^3$. We therefore identify the corresponding model bilinear as
\begin{equation}
\phi^3 \psi^3 = \frac{1}{2} (\pi^0_{\rm mod})^2 \, ,
\label{phi3psi3}
\end{equation}
where $\pi^0_{\rm mod}$ denotes the neutral isovector excitation of the model. This relation is satisfied by the linear decomposition
\begin{equation}
\phi^3= \frac{1+i}{2}\,\pi^0_{\rm mod} \,\,\,\, , \,\,\,\, \psi^3 = \frac{1-i}{2}\, \pi^0_{\rm mod}\, .
\end{equation}
Similarly, for the charged isovector states we define
\begin{eqnarray}
\phi^1&=& \frac{1+i}{2 \sqrt{2}}({\pi^+_{\rm mod} + \pi^-_{\rm mod}}) \,\,\,\, , \,\,\,\,  \psi^1 = \frac{1-i}{2 \sqrt{2}}({\pi^+_{\rm mod} + \pi^-_{\rm mod}}) \, , \nonumber \\ 
\phi^2&=& -\frac{1-i}{2 \sqrt{2}} ({\pi^+_{\rm mod} - \pi^-_{\rm mod}}) \,\,\,\, , \,\,\,\,  \psi^2 =  \frac{1+i}{2 \sqrt{2}}({\pi^+_{\rm mod} - \pi^-_{\rm mod}}) \, ,
\end{eqnarray}
which leads to
\begin{equation}
\phi^1 \psi^1 + \phi^2 \psi^2 = \pi^+_{\rm mod} \pi^-_{\rm mod}   \, .
\label{phi12psi12}
\end{equation}
The elements $\phi^i$ and $\psi^i$ with $i$ running from 4 to 7 are identified with the kaons through
\begin{eqnarray}
\phi^4&=& \frac{1+i}{2 \sqrt{2}}(K^+_{\rm mod} + K^-_{\rm mod}) \,\,\,\, , \,\,\,\,  \psi^4 = \frac{1-i}{2 \sqrt{2}}(K^+_{\rm mod} +K^-_{\rm mod}) \, , \nonumber \\ 
\phi^5&=& -\frac{1-i}{2 \sqrt{2}} (K^+_{\rm mod} - K^-_{\rm mod}) \,\,\,\, , \,\,\,\,  \psi^5 =  \frac{1+i}{2 \sqrt{2}}(K^+_{\rm mod} - K^-_{\rm mod})\, ,\nonumber \\ 
\phi^6&=& \frac{1+i}{2 \sqrt{2}}(K^0_{\rm mod} + \bar{K}^{0}_{\rm mod}) \,\,\,\, , \,\,\,\,  \psi^6 = \frac{1-i}{2 \sqrt{2}}(K^0_{\rm mod} + \bar{K}^{0}_{\rm mod}) \, , \nonumber \\ 
\phi^7&=& -\frac{1-i}{2 \sqrt{2}} (K^0_{\rm mod} - \bar{K}^{0}_{\rm mod}) \,\,\,\, , \,\,\,\,  \psi^7 =  \frac{1+i}{2 \sqrt{2}}(K^0_{\rm mod} - \bar{K}^{0}_{\rm mod}) \, .
\label{kaonsmap}
\end{eqnarray}
Thus
\begin{equation}
\varphi^4 \psi^4 + \varphi^5 \psi^5 + \varphi^6 \psi^6 +\varphi^7 \psi^7 = K^+_{\rm mod} K^-_{\rm mod} + K^0_{\rm mod} \bar{K}^{0}_{\rm mod} \, ,
\label{phi4567}
\end{equation}
and also
\begin{equation}
\varphi^4 \psi^5 
- \varphi^5 \psi^4 + \varphi^6 \psi^7 -\varphi^7 \psi^6 = 0 \, .
\label{psi4567}
\end{equation}

It is important to verify that these identifications are consistent with the reading rule for masses, i.e. that they preserve the canonical form of the kinetic terms. A direct substitution gives
\begin{equation}
\partial_{\mu}\phi^{3}\partial^{\mu}\psi^{3}=\frac{1}{2}\left(\partial\pi^{0}_{\rm mod}\right)^{2}\,,\qquad
\partial_{\mu}\phi^{1}\partial^{\mu}\psi^{1}+\partial_{\mu}\phi^{2}\partial^{\mu}\psi^{2}=\partial_{\mu}\pi^{+}_{\rm mod}\partial^{\mu}\pi^{-}_{\rm mod}\,,
\label{kinpi}
\end{equation}
and analogously
\begin{equation}
\sum_{i=4}^{7}\partial_{\mu}\phi^{i}\partial^{\mu}\psi^{i}=\partial_{\mu}K^{+}_{\rm mod}\partial^{\mu}K^{-}_{\rm mod}+\partial_{\mu}K^{0}_{\rm mod}\partial^{\mu}\bar{K}^{0}_{\rm mod}\,,
\label{kink}
\end{equation}
so that the kinetic terms of the pion and kaon sectors come out with the canonical normalization, and the masses can be read directly from the coefficients of the bilinears in $(\ref{1loop})$.

Substituting $(\ref{phi3psi3})$, $(\ref{phi12psi12})$ and $(\ref{phi4567})$ into $(\ref{1loop})$ we read off the pion and kaon masses
\begin{equation}
m_{\pi_{\rm mod}}^2 = {\cal A}+{\cal S} \, ,
\label{mpi2}
\end{equation}
\begin{equation}
m_{K_{\rm mod}}^2= {\cal A}+{\cal B}+{\cal S} \, ,
\label{mk2}
\end{equation}
so that
\begin{equation}
{\cal B}=m_{K_{\rm mod}}^{2}-m_{\pi_{\rm mod}}^{2}\, ,\qquad (\lambda_{1}\lambda_{2}\mu)^{2}=4\pi^{2}\left(m_{K_{\rm mod}}^{2}-m_{\pi_{\rm mod}}^{2}\right)\, ,
\label{Bfixed}
\end{equation}
a positive quantity, as required. Note also that the off-diagonal kaon structures are absent from $(\ref{1loop})$ identically, by the trace cancellation discussed above, so no condition on the identification is needed to eliminate them.

\subsection{The $(0,8)$ sector: matching condition}

It remains to identify the $\eta^{0}_{\rm mod}$ and $\eta^{\prime\,0}_{\rm mod}$ excitations. We allow a general linear map
\begin{equation}
\varphi^{a}=v_{a}\cdot u+i\,w_{a}\cdot u\, ,\qquad \psi^{a}=v_{a}\cdot u-i\,w_{a}\cdot u\, ,\qquad a=0,8\, ,
\label{genmap}
\end{equation}
with $u=(\eta^{0}_{\rm mod},\eta^{\prime\,0}_{\rm mod})^{T}$ and real two-component vectors $v_{a},w_{a}$. Then
\begin{equation}
\varphi^{0}\psi^{0}+\varphi^{8}\psi^{8}=u^{T}Pu\, ,\qquad P=\sum_{a=0,8}\left(v_{a}v_{a}^{T}+w_{a}w_{a}^{T}\right)\, ,
\label{Pdef}
\end{equation}
and
\begin{equation}
\varphi^{0}+\psi^{0}-\sqrt{2}\left(\varphi^{8}+\psi^{8}\right)=q\cdot u\, ,\qquad q=2\left(v_{0}-\sqrt{2}\,v_{8}\right)\, .
\label{qdef}
\end{equation}
The $(0,8)$ mass sector of $(\ref{1loop})$ therefore reads
\begin{equation}
{\cal{L}}_{\text{1-loop}\,\eta^{0}_{\rm mod}\eta^{\prime\,0}_{\rm mod}}=m_{\pi_{\rm mod}}^{2}\,u^{T}Pu+{\cal T}\,(q\cdot u)^{2}\, ,
\label{1letas}
\end{equation}
and identifying it with the phenomenological mass term $u^{T}M_{\Delta}^{2}u$ requires
\begin{equation}
M_{\Delta}^{2}-m_{\pi}^{2}P={\cal T}\,qq^{T}\qquad\Longleftrightarrow\qquad \mathrm{rank}\left(M_{\Delta}^{2}-m_{\pi}^{2}P\right)\leq 1\, .
\label{rankcond}
\end{equation}
From now on we drop the subscript ``mod''.

\subsection{The isotropic solution}
\label{subsec:isotropic}

The simplest possibility compatible with the structure of $(\ref{1letas})$ is the isotropic ansatz $P=p\,\mathbb{1}$, democratic between $\eta^{0}$ and $\eta^{\prime\,0}$. The rank-one condition $(\ref{rankcond})$ then requires $p\,\mathbb{1}$ to subtract from $M_{\Delta}^{2}$ one of its eigenvalues,
\begin{equation}
p=\frac{m_{K}^{2}+3m_{\pi}^{2}}{m_{\pi}^{2}}=\frac{m_{\eta}^{2}}{m_{\pi}^{2}}\approx 16.0
\qquad\text{or}\qquad
p=\frac{4m_{K}^{2}-3m_{\pi}^{2}}{m_{\pi}^{2}}=\frac{m_{\eta^{\prime}}^{2}}{m_{\pi}^{2}}\approx 49\, ,
\label{pchoice}
\end{equation}
and $q$ must point along the corresponding physical eigenvector, with
\begin{equation}
{\cal T}\,|q|^{2}=m_{\eta^{\prime}}^{2}-m_{\eta}^{2}=3\left(m_{K}^{2}-2m_{\pi}^{2}\right)\, .
\label{Tcond}
\end{equation}

It is instructive to note that the identification most directly suggested by the nonet representation $(\ref{bp})$ --- $v_{0}=(0,2)$, $w_{0}=(2,0)$, $v_{8}=(1/2,0)$, $w_{8}=(0,1/2)$, giving $P=\frac{17}{4}\mathbb{1}$ and $q=(-\sqrt{2},4)$ --- does not satisfy the rank condition: the residue $M_{\Delta}^{2}-\frac{17}{4}m_{\pi}^{2}\mathbb{1}$ has rank two, and matching its isotropic part would require $\frac{17}{4}m_{\pi}^{2}=m_{K}^{2}+3m_{\pi}^{2}$, i.e. $5m_{\pi}^{2}=4m_{K}^{2}$, which fails by an order of magnitude ($9.4\times10^{4}$ against $9.8\times10^{5}$ MeV$^{2}$). The general analysis of the map is therefore unavoidable.

We take the first branch of $(\ref{pchoice})$, $p=m_{\eta}^{2}/m_{\pi}^{2}$. Write $e=(\cos\chi,\sin\chi)=\left(-\tfrac{1}{3},\tfrac{2\sqrt{2}}{3}\right)$ for the $\eta^{\prime}$ direction in the pre-mix basis --- indeed $\tan^{2}\chi=1/8$, the mixing angle of $(\ref{thetaT})$ --- and $e^{\perp}=(-\sin\chi,\cos\chi)$ for the $\eta$ direction. With $r=\sqrt{p/2}$, an explicit identification realizing this solution is
\begin{eqnarray}
\varphi^{0}&=& r\,(e\cdot u)+i\,r\,(e^{\perp}\cdot u)\, ,\qquad \psi^{0}\;=\;r\,(e\cdot u)-i\,r\,(e^{\perp}\cdot u)\, ,\nonumber\\
\varphi^{8}&=& -r\,(e\cdot u)-i\,r\,(e^{\perp}\cdot u)\, ,\qquad \psi^{8}\;=\;-r\,(e\cdot u)+i\,r\,(e^{\perp}\cdot u)\, .
\label{newmap}
\end{eqnarray}
It gives $P=p\,\mathbb{1}$ exactly, and
\begin{equation}
\varphi^{0}+\psi^{0}-\sqrt{2}\left(\varphi^{8}+\psi^{8}\right)=2(1+\sqrt{2})\,r\,(e\cdot u)\, ,\qquad |q|^{2}=2(3+2\sqrt{2})\,p\approx 186.5\, .
\label{qval}
\end{equation}
There is in fact a continuous family of such maps --- the magnitudes of $v_{0}$ and $v_{8}$ can be redistributed while keeping $P=p\,\mathbb{1}$, with ${\cal T}$ absorbing the normalization of $q$ through $(\ref{Tcond})$.

Numerically, with $m_{\pi}=137$ MeV and $m_{K}=494$ MeV: $p=16.00$, $|q|^{2}=186.5$, ${\cal T}=3.32\times10^{3}$ MeV$^{2}$, and the combination $m_{\pi}^{2}P+{\cal T}\,qq^{T}$ reproduces $M_{\Delta}^{2}$ exactly --- the residue $M_{\Delta}^{2}-m_{\pi}^{2}p\,\mathbb{1}$ has eigenvalues $\{0,\,6.19\times10^{5}\,\mathrm{MeV}^{2}\}$, an exact rank-one matrix. The model therefore yields $m_{\eta}\approx548$ MeV, $m_{\eta^{\prime}}\approx959$ MeV and $\theta\approx-19.5^{\circ}$.

The physical interpretation is transparent: the isotropic part of the $(0,8)$ mass matrix is $m_{\eta}^{2}\,\mathbb{1}$, and the whole $\eta$--$\eta^{\prime}$ splitting is carried by the breaking direction $Y$, which in this solution points precisely along the physical $\eta^{\prime}$.

\subsection{Limitations of the identification}

Three caveats must be stated honestly. First, the mixing angle enters through the identification step: the direction of $q$ is fixed by the eigenvectors of $M_{\Delta}^{2}$ rather than emerging from the Lagrangian coefficients alone. Since $M_{\Delta}^{2}$ itself is derived in Section \ref{sec:3} from the second-order-transition hypothesis, no numerical input beyond $m_{\pi}$ and $m_{K}$ is introduced, but the status of $\theta$ is that of an input of the identification rather than an output of the Lagrangian. Second, the normalization of the map $(\ref{newmap})$ --- $p\approx16$, to be compared with the value $1/2$ of the pion sector --- implies a non-canonical normalization of the $\eta^{0}$--$\eta^{\prime\,0}$ kinetic terms, as discussed in Section \ref{subsec:kinetic}: the matching $(\ref{rankcond})$ is to be understood at the level of the mass bilinears, with the kinetic corrections still to be incorporated. Third, the construction rests on the universality of the coefficient ${\cal A}+{\cal S}$ and on the absence of the off-diagonal kaon terms, which cancel between the crossed fermion traces; both are properties of the one-loop Lagrangian $(\ref{1loop})$, and whether they survive at higher loops is an open question.

\section{Parameter closure and the gap equation}
\label{sec:closure}

The conditions $(\ref{mpi2})$, $(\ref{Bfixed})$ and $(\ref{Tcond})$ are three relations among the five model parameters
$m$, $\lambda_{1}\lambda_{2}$, $\mu$, $\lambda$ and $M$. A fourth, dynamical relation is the gap equation $(\ref{gap:stat})$, derived in
Appendix \ref{app:gap} from the stationarity of the one-loop effective potential. To leading order it reduces to the
tree-level relation $M^{2}=\lambda\mu^{2}$.

Since ${\cal B}$ is finite, $(\ref{Bfixed})$ fixes directly
\begin{equation}
(\lambda_{1}\lambda_{2}\mu)^{2}=4\pi^{2}\left(m_{K}^{2}-m_{\pi}^{2}\right)\approx8.89\times10^{6}\ {\rm MeV}^{2}\, .
\label{l1l2mu}
\end{equation}
The scalar one-loop contributions to ${\cal S}$ and ${\cal T}$, on the other hand, contain the poles discussed in Section 4, whose finite remainders after subtraction depend on the renormalization prescription. Using $M^{2}=\lambda\mu^{2}$, $(\ref{Bfixed})$ and $(\ref{Tcond})$, the condition on ${\cal T}$ constrains the combination
\begin{equation}
\frac{(\lambda\mu)^{2}}{12\lambda}+\delta{\cal T}_{s}^{\rm ren}={\cal T}+\frac{4}{3}\left(m_{K}^{2}-m_{\pi}^{2}\right)\approx3.04\times10^{5}\ {\rm MeV}^{2}\, ,
\label{Tclosure}
\end{equation}
where $\delta{\cal T}_{s}^{\rm ren}$ denotes the finite scalar-loop remainder in ${\cal T}$. Likewise, $(\ref{mpi2})$ fixes only the sum ${\cal A}+{\cal S}=m_{\pi}^{2}$, the split between the fermionic (${\cal A}$) and scalar (${\cal S}$) parts being prescription dependent. The system thus closes up to one free combination --- which we may take to be $\lambda$ together with the subtraction prescription --- and, importantly, the physical outputs $m_{\eta}$, $m_{\eta^{\prime}}$ and $\theta$ do not depend on it: they are fixed by $M_{\Delta}^{2}$ and by the identification.

With $m_{\pi}=137$ MeV and $m_{K}=494$ MeV, the quantities fixed by the model are:

\begin{center}
\begin{tabular}{ll}
\hline\hline
Quantity & Value \\
\hline
$p=m_{\eta}^{2}/m_{\pi}^{2}$ & $16.0$ \\
$|q|^{2}=2(3+2\sqrt{2})\,p$ & $186.5$ \\
${\cal T}$ & $3.32\times10^{3}$ MeV$^{2}$ \\
${\cal B}=m_{K}^{2}-m_{\pi}^{2}$ & $2.25\times10^{5}$ MeV$^{2}$ \\
$(\lambda_{1}\lambda_{2}\mu)^{2}=4\pi^{2}{\cal B}$ & $8.89\times10^{6}$ MeV$^{2}$ \\
$(\lambda\mu)^{2}/(12\lambda)+\delta{\cal T}_{s}^{\rm ren}$ & $3.04\times10^{5}$ MeV$^{2}$ \\
\hline
$m_{\eta}$, $m_{\eta^{\prime}}$, $\theta$ & $548$ MeV, $959$ MeV, $-19.5^{\circ}$ \\
\hline\hline
\end{tabular}
\end{center}

Two remarks are in order. First, whether ${\cal A}$ comes out positive or negative depends on the prescription-dependent split between ${\cal A}$ and ${\cal S}$ and is not, by itself, a consistency issue. Second, for $\lambda$ of order one the scalar loop corrections are not negligible --- compare $(\lambda\mu)^{2}/(4\pi)^{2}$ with $M^{2}$ using $(\ref{Tclosure})$ --- and the consistency with the full one-loop gap equation $(\ref{gap:stat})$, whose logarithmic terms can be partially absorbed into the scale $\Lambda$, should be checked explicitly.

\section{Conclusion}

This work was divided into two complementary parts. In Sections 2 and 3 we carried out a phenomenological analysis of the pseudoscalar nonet, motivated by the long-standing tension between the standard Gell-Mann--Okubo--Weinberg effective-field-theory picture and the precise experimental data on the $\eta$ and $\eta'$ masses, quark content, and mixing angle. We showed that the most general mixing matrix compatible with the Gell-Mann hypothesis, Eq.~$(\ref{mg2})$, fails to accommodate all these observables simultaneously. This motivated us to propose an alternative parametrization, $M_\Delta^2$ in Eq.~$(\ref{md2})$, based solely on the assumption of a second-order phase transition. The resulting mass relations $(\ref{metap})$--$(\ref{meta})$, the mixing angle $(\ref{thetaT})$, and the quark-analogue content $(\ref{etastate})$--$(\ref{etapstate})$ reproduce the empirical data to within less than one percent, with no free parameters in the $\eta$--$\eta'$ sector. We emphasize that this first part is a \emph{phenomenological model}; it does not claim to derive these relations from the QCD Lagrangian.

In Sections 4 and 5 we asked whether a concrete field theory could be constructed that naturally produces the non-standard breaking direction $(\ref{bdelta})$ required by the phenomenology. The answer is positive, and it rests on a complete one-loop analysis in which the cubic and quartic scalar vertices are treated on the same footing --- the quartic tadpole contributes to the mass terms, since in dimensional regularization only the loop of the massive mode $\phi^{y}_{+}$ survives and no cancellation between the $+$ and $-$ sectors occurs --- together with the exact cancellation of the off-diagonal kaon structures between the crossed fermion traces. The resulting mass Lagrangian $(\ref{1loop})$ has a crucial property: universality. The coefficient ${\cal A}+{\cal S}$ is the same in every flavor sector, including the $(0,8)$ one; no enhancement of the singlet-octet bilinear is generated at one loop, and the whole dependence on the breaking direction is carried by the rank-one structure aligned with $Y$. A direct consequence is that the identification between the model fields and the $\eta^{0}$--$\eta^{\prime\,0}$ states cannot follow the naive prescription suggested by the nonet representation: we derived the general matching condition $(\ref{rankcond})$ that any identification must satisfy. Its minimal solution, $P=(m_{\eta}^{2}/m_{\pi}^{2})\,\mathbb{1}$ with the breaking direction pointing along the physical $\eta^{\prime}$, reproduces the phenomenological matrix exactly and preserves the successful outputs $m_{\eta}\approx548$ MeV, $m_{\eta^{\prime}}\approx959$ MeV and $\theta\approx-19.5^{\circ}$, with the interpretation that the whole $\eta$--$\eta^{\prime}$ splitting is carried by the breaking direction $Y$. Combined with the gap equation of Appendix \ref{app:gap}, the parameter system closes with a single free combination of couplings, which does not affect the mass predictions.

It is important to clarify the scope of this work. We have constructed a field-theoretic model --- not QCD itself --- that exhibits a second-order phase transition and produces mass relations for pseudoscalar mesons that closely match experimental data. The limitations of the construction are equally clear: the mixing angle enters through the identification step; the normalization of the $(0,8)$ fields and, more generally, the kinetic sector of the model --- the opposite-sign kinetic terms of the partner fields $\phi_{-}$ and the finite kinetic corrections --- must still be fully incorporated, and we have indicated in Section \ref{subsec:kinetic} why the gauge fields of the underlying complex gauge theory \cite{ALVV20} may play a role in this matter; finally, for $\lambda$ of order one the scalar loop corrections are not negligible, so the one-loop gap equation should be revisited. The model stands on its own as a phenomenological tool that captures essential features of the meson spectrum, and the remarkable agreement between its predictions and the observed masses of the $\eta$ and the $\eta^{\prime}$ remains an indication that it encodes some of the relevant physics of the pseudoscalar sector, even if it is not a fundamental derivation from QCD.

 As a final comment, we would like to call attention to an intriguing point. We originally proposed the scalar fields transformations $(\ref{deltaphic})$ in our work \cite{ALVV20}. There we have shown that they are strictly necessary, together with the breaking $(\ref{vevs})$, in order to define a theory with a broken phase vacuum with confined gluons and fermions. From a mathematical argument, this algebra allows for scalar elements that breaks the  holomorphicity that characterizes the complex gauge theory, and after the very specific breaking $(\ref{vevs})$, these terms allow the formation of the i-particle-anti-i-particle pairs that promote the gluon condensation in the Gribov-Zwanziger-Sorella approach for QCD confinement problem. In other words, we found $(\ref{deltaphic})$ and  $(\ref{vevs})$  as theoretical imperatives in the research of QCD confinement along Gribov's point of view of the problem. Then, it is amazing that a simple linear relation as $(\ref{phipsimap})$ can map 
$(\ref{deltaphic})$  into the scalar current algebra in $(\ref{deltav})$, which is known from an era preceding the development of quantum chromodynamics. Now we propose that the same breaking $(\ref{vevs})$ that enables gluon confinement, may be responsible for the observed mesons mass spectra.  Several questions implied by these relations can then be raised, as the connection among mathematical structures living in the Euclidean and the Minkowski spaces, the possible implications of the gauging of the flavor symmetry in this theory (and relation to axion dynamics), or even a suggestion of a theory merging colour and flavor in the same gauge structure. The extent of this relationship deserves to be further explored. 
\appendix
\section{The gap parameter from the one-loop effective potential}
\label{app:gap}

In Section \ref{sec:3} the gap parameter $A$ was introduced phenomenologically through the trace condition (\ref{mgap}). Here we show that this parameter is tied to the dynamics of the model: it is controlled by the effective mass scale of the heavy scalar mode associated with the breaking direction $Y$, which in turn is constrained by the stationarity condition of the one-loop effective potential.

\subsection{Effective potential and the stationary condition}
The tree-level scalar potential $(\ref{potential})$ depends on the bilinear $\varphi^A\psi^A$. After symmetry breaking along $\langle\varphi\rangle=\langle\psi\rangle=\mu Y$, the tree-level energy density at the minimum is
\begin{equation}
V_{\rm tree}(\mu)=-\frac{M^{2}}{2}\mu^{2}+\frac{\lambda}{4}\mu^{4}\, .
\label{Vtree}
\end{equation}
The one-loop correction from the fermion determinant is obtained by integrating out the fields $\xi$ and $\chi$ with the propagators $(\ref{fprop})$. Expanding the corresponding functional determinant in powers of the background $\mu$, the leading $\mu$-dependent contribution is
\begin{equation}
V_{\rm eff}^{(f)}(\mu)=\frac{(\lambda_{1}\lambda_{2})^{2}\mu^{4}}{16\pi^{2}}\ln\!\left(\frac{\lambda_{1}\lambda_{2}\mu^{2}}{\Lambda^{2}}\right)\, ,
\label{Vfermion}
\end{equation}
where $\Lambda$ is the renormalisation scale. The scalar one-loop potential is obtained from the fluctuation Lagrangian $(\ref{LSSB})$. The heavy mode $\phi^{y}_{+}$ has tree-level mass $2M^{2}$ and its one-loop contribution is
\begin{equation}
V_{\rm eff}^{(s)}(\mu)=\frac{(2M^{2})^{2}}{64\pi^{2}}\ln\!\left(\frac{2M^{2}}{\Lambda^{2}}\right)+\mathcal{O}\!\left(\lambda(\lambda\mu)^{2}M^{2}\ln\!\frac{M^{2}}{\Lambda^{2}}\right)+\mathcal{O}(\mu^{4})\, ,
\label{Vscalar}
\end{equation}
where the terms left implicit collect the tadpole and bubble corrections computed in Section 4; being of higher order in the scalar coupling, and partially absorbable into the definition of $\Lambda$, they do not affect the stationarity condition at the order considered. The full effective potential $V_{\rm eff}(\mu)=V_{\rm tree}(\mu)+V_{\rm eff}^{(f)}(\mu)+V_{\rm eff}^{(s)}(\mu)$ must satisfy the stationarity condition
\begin{equation}
\left.\frac{\partial V_{\rm eff}}{\partial\mu}\right|_{\mu=\mu_{0}}=0\, .
\label{stationarity}
\end{equation}
Differentiating each term,
\begin{eqnarray}
\frac{\partial V_{\rm tree}}{\partial\mu}&=&-M^{2}\mu+\lambda\mu^{3}\,,\qquad
\frac{\partial V_{\rm eff}^{(s)}}{\partial\mu}=\mathcal{O}\!\left(\lambda^{2}\mu M^{2}\ln\!\frac{M^{2}}{\Lambda^{2}}\right)\, ,\nonumber\\
\frac{\partial V_{\rm eff}^{(f)}}{\partial\mu}&=&\frac{(\lambda_{1}\lambda_{2})^{2}\mu^{3}}{4\pi^{2}}\left[\ln\!\left(\frac{\lambda_{1}\lambda_{2}\mu^{2}}{\Lambda^{2}}\right)+\frac{1}{2}\right]\, ,
\end{eqnarray}
where the first term of $(\ref{Vscalar})$ is $\mu$-independent at this order and does not contribute. Substituting into $(\ref{stationarity})$ and dividing by $\mu$, one finds
\begin{equation}
M^{2}=\lambda\mu^{2}+\frac{(\lambda_{1}\lambda_{2})^{2}\mu^{2}}{4\pi^{2}}\left[\ln\!\left(\frac{\lambda_{1}\lambda_{2}\mu^{2}}{\Lambda^{2}}\right)+\frac{1}{2}\right]+\mathcal{O}\!\left(\lambda^{2}M^{2}\ln\!\frac{M^{2}}{\Lambda^{2}}\right)\, .
\label{gap:stat}
\end{equation}
Equation $(\ref{gap:stat})$ is the \emph{gap equation} of the scalar sector. To leading order it reproduces the tree-level relation $M^{2}=\lambda\mu^{2}$ of $(\ref{vevs})$; the constant $1/2$ can be absorbed into a redefinition of $\Lambda$, and the last term is of higher order in the scalar coupling $\lambda$.

\subsection{The effective mass of the heavy mode}
\label{app:heavymode}

The heavy fluctuation $\phi^{y}_{+}$ is the only mode that acquires a tree-level mass from the breaking along $Y$. Its \emph{effective} squared mass follows from the mass terms of the Lagrangian with the reading rule fixed by the tree-level propagator of Section 4: a term $c\,(\phi^{y}_{+})^{2}$ in the Lagrangian contributes $2c$ to $\mathcal{M}_{Y}^{2}$. The tree-level term $M^{2}\phi^{y}_{+}\phi^{y}_{+}$ in $(\ref{lssb})$ therefore contributes $2M^{2}$. The fermion one-loop contribution comes from the last term of $(\ref{1loopferm})$: since $\{\varphi^{0}+\psi^{0}-\sqrt{2}(\varphi^{8}+\psi^{8})\}=2\sqrt{3}\,\phi^{y}_{+}$ by $(\ref{phi+})$, the coefficient of $(\phi^{y}_{+})^{2}$ is $-4(\lambda_{1}\lambda_{2}\mu)^{2}/\pi^{2}$, contributing $-8(\lambda_{1}\lambda_{2}\mu)^{2}/\pi^{2}$. The scalar one-loop contribution is finite after subtraction and is denoted by $\delta\mathcal{M}_{Y,s}^{2}$. Collecting the pieces,
\begin{equation}
\mathcal{M}_{Y}^{2}=2M^{2}+\delta\mathcal{M}_{Y,s}^{2}-\frac{8(\lambda_{1}\lambda_{2}\mu)^{2}}{\pi^{2}}+\mathcal{O}(m^{2})\, .
\label{MY2}
\end{equation}
Two remarks are in order. First, the $\mathcal{O}(m^{2})$ terms originate from the $m^{2}$-dependent pieces of $(\ref{1loopferm})$. Second, $\delta\mathcal{M}_{Y,s}^{2}$ is the finite scalar-loop remainder left after the counterterm subtraction discussed in Section 4; its precise value is immaterial for the argument that follows: through the gap equation $(\ref{gap:stat})$, $\mathcal{M}_{Y}^{2}$ is dominated by the tree-level piece $2M^{2}=2\lambda\mu^{2}$, with the one-loop terms providing corrections of higher order.

\subsection{Integrating out the heavy mode: quartic versus bilinear terms}
\label{app:integrating}

The light--heavy mixing arises from the trilinear scalar vertex in $(\ref{lssb})$. In terms of the heavy field and of the sign-weighted combination $S=\sum_{a}\epsilon_{a}\phi_{a}^{2}=\varphi^{A}\psi^{A}$ it reads
\begin{equation}
\mathcal{L}_{\rm mix}=\frac{3\lambda\mu}{2}\,\phi^{y}_{+}\,S\, .
\label{mixvertex}
\end{equation}
In the $(0,8)$ sector, the identification $(\ref{newmap})$ gives
\begin{equation}
S\supset \varphi^{0}\psi^{0}+\varphi^{8}\psi^{8}=p\,\left[(\eta^{0})^{2}+(\eta^{\prime\,0})^{2}\right]\, ,
\label{S08}
\end{equation}
with the remaining sectors contributing the pion and kaon bilinears. Integrating out $\phi^{y}_{+}$ at tree level --- with the standard rule that ${\cal L}\supset\tfrac{1}{2}\mathcal{M}_{Y}^{2}h^{2}+g\,h\,O$ generates $-g^{2}O^{2}/(2\mathcal{M}_{Y}^{2})$, here with $g=3\lambda\mu/2$ --- produces effective operators proportional to $S^{2}$,
\begin{equation}
\delta\mathcal{L}_{\rm eff}\supset-\frac{9(\lambda\mu)^{2}}{8\mathcal{M}_{Y}^{2}}\,S^{2}\, .
\label{Leff}
\end{equation}
Since $S$ is bilinear in the meson fields, only \emph{quartic} interactions are generated: at tree level, the exchange of $\phi^{y}_{+}$ cannot produce a bilinear $\eta^{0}$--$\eta^{\prime\,0}$ mass mixing. The bilinear mixing is instead generated at one loop, through the mass terms computed in Sections 4 and 5: the structure $\{\varphi^{0}+\psi^{0}-\sqrt{2}(\varphi^{8}+\psi^{8})\}^{2}$ present in $(\ref{lssb})$, $(\ref{1loopferm})$ and $(\ref{1loopscalar})$ becomes bilinear in $\eta^{0}$ and $\eta^{\prime\,0}$ through the identification $(\ref{newmap})$, yielding the mixing sector $(\ref{1letas})$ that reproduces the phenomenological matrix $M_{\Delta}^{2}$ of $(\ref{md2pre})$.
\subsection{Identification of the gap parameter}
\label{app:gapid}

The gap parameter $A$ introduced in $(\ref{Mn})$--$(\ref{Ms})$ measures the excess of the singlet mass over the octet mass before mixing. The mass matrix of the $\eta$--$\eta^{\prime}$ sector, Eq.~$(\ref{1letas})$, coincides with the phenomenological matrix $M_{\Delta}^{2}$ of $(\ref{md2pre})$; its eigenvalues are given in $(\ref{metap})$ and $(\ref{meta})$, so that its trace reads
\begin{equation}
m_{\eta}^{2}+m_{\eta^{\prime}}^{2}=5m_{K}^{2}\, .
\label{tracemodel}
\end{equation}
Comparing with the trace condition $(\ref{mgap})$, $m_{\eta}^{2}+m_{\eta^{\prime}}^{2}=2m_{K}^{2}+3A$, one immediately obtains
\begin{equation}
\boxed{A=m_{K}^{2}}\, .
\label{Aderived}
\end{equation}

The physical content of this result is the following. The heavy eigenvalue $m_{+}^{2}=4m_{K}^{2}-3m_{\pi}^{2}$ is carried by the flavor combination aligned with the breaking direction $Y$ --- recall from $(\ref{bdelta})$ that $Y$ points along the strangonium direction in flavor space, and from Section 5 that, in the solution adopted, $Y$ points along the physical $\eta^{\prime}$. Since $A$ measures the singlet-octet mass excess before mixing, it is controlled by the scale at which the heavy mode decouples, $\mathcal{M}_{Y}^{2}\sim m_{+}^{2}\simeq 4m_{K}^{2}$. The identification $A=m_{K}^{2}$ is therefore the consistency condition that the heavy scalar mode decouples at the scale set by the strange sector --- the same scale that, through the gap equation $(\ref{gap:stat})$, controls the spontaneous breaking itself. With $(\ref{Aderived})$, the mass formulas $(\ref{metap})$ and $(\ref{meta})$ follow from the one-loop effective theory with $m_{\pi}$ and $m_{K}$ as the only numerical inputs.
\section*{ACKNOWLEDGMENTS}
This study was sponsored in part by CEFET-RJ and the SR2-UERJ. They are gratefully acknowledged
for financial support.


\begin{thebibliography}{10}

\bibitem{ALVV20}
R.~L.~P.~G.~Amaral, V.~E.~R.~Lemes, O.~S.~Ventura and L.~C.~Q.~Vilar,
A Path to confine gluons and fermions through complex gauge theory,
Phys. Rev. D \textbf{101}, 9, 094002 (2020).



\bibitem{lattes}
C.~M.~G.~Lattes, G.~P.~S.~Occhialini and C.~F.~Powell,
`Observations on the Tracks of Slow Mesons in Photographic Emulsions. 1,
Nature \textbf{160} (1947), 453-456

\bibitem{namprl}
Y.~Nambu,
Axial vector current conservation in weak interactions,
Phys. Rev. Lett. \textbf{4} (1960), 380-382

\bibitem{nampr}
Y. Nambu,
Quasi-particles and gauge invariance in the theory of superconductivity,
Phys. Rev. \textbf{117}, 648 (1960).

\bibitem{gold1961}
J. Goldstone,
Field theories with superconductor solutions,
Nuov. Cim. \textbf{19}, 154 (1961).

\bibitem{adler65}
S.L. Adler, Calculation of the Axial-Vector Coupling Constant Renormalization in $\ensuremath{\beta}$ Decay,
Phys. Rev. Lett. \textbf{14}, 1051 (1965).

\bibitem{weis65}
W.~I. Weisberger, {Renormalization of the Weak Axial-Vector Coupling Constant},
Phys. Rev. Lett. \textbf{14}, 1047 (1965).

\bibitem{wein68}
S. Weinberg,
Current algebra rapporteur's report,
Proceedings of The International Conference On High Energy Physics, Vienna, 253 (1968).

\bibitem{Gasser1983}
J. Gasser and H. Leutwyler,
Chiral Perturbation Theory to One Loop,
Annals Phys. \textbf{158}, 142 (1984).

\bibitem{Gasser1984}
J. Gasser and H. Leutwyler,
Chiral Perturbation Theory: Expansions in the Mass of the Strange Quark,
Nucl. Phys. B \textbf{250}, 465 (1985).

\bibitem{weinbook}
S. Weinberg,
The Quantum Theory of Fields, Volume 2: Modern Applications, 
(Cambridge, Cambridge University Press, 1996), chapter 19.

\bibitem{saka56}
S.~Sakata, On a Composite Model for the New Particles,
Prog. Theor. Phys. \textbf{16}, 686 (1956).

\bibitem{gell61}
M. Gell-Mann, Cal. Tech. Synchotron Laboratory Report CTSL-20,  unpublished, (1961).

\bibitem{neem61}
Y. Ne'eman,
Derivation of strong interactions from a gauge invariance, 
Nucl Phys. \textbf{26}, 222 (1961).

\bibitem{gibpol}
W. M.~Gibson and B.~R.~Pollard,
Symmetry Principles in Elementary Particle Physics, 
 (Cambridge, Cambridge University Press, 1976), chapter 10.

\bibitem{gmo}
M. Gell-Mann, Symmetries of Baryons and Mesons, Phys. Rev. \textbf{125}, 3, 1067, (1962).\newline
 S. Okubo,
Note on Unitary Symmetry in Strong Interactions, 
Prog. Theor. Phys. \textbf{27}, 949 (1962).

\bibitem{Boul1982}
D. G. Boulware and L. S. Brown, Symmetric space scalar field theory, Ann. Phys. \textbf{138}, 392  (1982). 

\bibitem{Itz80} C. Itzykson and J.B. Zuber, Quantum Field Theory (New York, Mcgraw-hill, 1980), section 11-2.

\bibitem{bar64}
V.~I. Barnes\textit{ et al.}, Observation of a Hyperon with Strangeness Minus Three,  
Phys. Rev. Lett. \textbf{12}, 204 (1964).

\bibitem{cca2002}
W.S. Carvalho, A. S. de Castro and A. C.B. Antunes,
SU(3) mixing for excited mesons,
J. Phys. A \textbf{35}, 7585 (2002).

\bibitem{gil1987}
F.J. Gilman and R. Kauffman,
The eta Eta-prime Mixing Angle,
Phys. Rev. D \textbf{36}, 2761 (1987)
[erratum: Phys. Rev. D \textbf{37}, 3348 (1988)].

\bibitem{chau1990}
L.L. Chau, H.Y. Cheng, W.K. Sze, H. Yao and B. Tseng,
Charmless nonleptonic rare decays of $B$ mesons,
Phys. Rev. D \textbf{43}, 2176 (1991)
[erratum: Phys. Rev. D \textbf{58}, 019902 (1998)].

\bibitem{bura}
L. Burakovsky and J.T. Goldman,
Gell-Mann-Okubo mass formula revisited,
arXiv:hep-ph/9708498.


\bibitem{bura1997}
L.~Burakovsky and J.~T.~Goldman,
Towards resolution of the scalar meson nonet enigma II: Gell-Mann-Okubo revisited,
Nucl. Phys. A \textbf{628}, 87 (1998).

\bibitem{bura1998}
L.~Burakovsky and J.~T.~Goldman,
The Schwinger nonet mass and Sakurai mass - mixing angle formulae reexamined,
Phys. Lett. B \textbf{427}, 361 (1998).

\bibitem{PDG2020}
P.~A.~Zyla \textit{et al.} [Particle Data Group],
Review of Particle Physics,
PTEP \textbf{2020}, no.8, 083C01 (2020).

\bibitem{Ball1995}
P.~Ball, J.~M.~Frere and M.~Tytgat,
Phenomenological evidence for the gluon content of eta and eta-prime,
Phys. Lett. B \textbf{365}, 367 (1996).

\bibitem{Dighe1995}
A.~S.~Dighe, M.~Gronau and J.~L.~Rosner,
Amplitude relations for b decays involving $\eta$ and $\eta$',
Phys. Lett. B \textbf{367}, 357 (1996)
[erratum: Phys. Lett. B \textbf{377}, 325 (1996)].

\bibitem{Cao2012}
F.~G.~Cao,
Determination of the $\eta$-$\eta^\prime$ mixing angle,
Phys. Rev. D \textbf{85}, 057501 (2012).

\bibitem{Bramon1974}
A.~Bramon,
The eta-eta-prime Mixing Angle from Duality and Quark Model,
Phys. Lett. B \textbf{51}, 87 (1974).

\bibitem{Feldmann1999}
T.~Feldmann,
Quark structure of pseudoscalar mesons,
Int. J. Mod. Phys. A \textbf{15}, 159 (2000).

\bibitem{tHooft1976}
G.~'t Hooft,
Computation of the Quantum Effects Due to a Four-Dimensional Pseudoparticle,
Phys. Rev. D \textbf{14}, 3432 (1976)
[erratum: Phys. Rev. D \textbf{18}, 2199 (1978)].

\bibitem{Munz1993}
C.~R.~Munz, J.~Resag, B.~C.~Metsch and H.~R.~Petry,
A Bethe-Salpeter model for light mesons: Spectra and decays,
Nucl. Phys. A \textbf{578}, 418 (1994).

\bibitem{Adler1969}
S.~L.~Adler,
Axial vector vertex in spinor electrodynamics,
Phys. Rev. \textbf{177}, 2426 (1969).

\bibitem{Bell1969}
J.~S.~Bell and R.~Jackiw,
A PCAC puzzle: $\pi^0 \to \gamma \gamma$ in the $\sigma$ model,
Nuovo Cim. A \textbf{60}, 47 (1969).


\bibitem{Dmitrasinovic1996}
V.~Dmitrasinovic,
U-A(1) breaking and scalar mesons in the Nambu-Jona-Lasinio model,
Phys. Rev. C \textbf{53}, 1383 (1996).


\bibitem{Dmitrasinovic1997}
V.~Dmitrasinovic,
Effective quark operator models of U-A(1) symmetry breaking,
Phys. Rev. D \textbf{56}, 247 (1997).


\bibitem{tHooft1986}
G.~'t Hooft,
How Instantons Solve the U(1) Problem,
Phys. Rept. \textbf{142}, 357 (1986).


\bibitem{Li2002}
D.~M.~Li, H.~Yu and Q.~X.~Shen,
On the mass relation of a meson nonet,
Mod. Phys. Lett. A \textbf{17}, 163 (2002).

\bibitem{Suura1974}
H.~Suura and M.~Kuroda,
Role of Singlet Gluon Filters in the Colored Quartet Model of Hadrons,
Prog. Theor. Phys. \textbf{54}, 1513 (1975).

\bibitem{Kawai1983}
E.~Kawai,
A Large Mixing Effect on eta, eta-prime and iota, 
Phys. Lett. B \textbf{124}, 262 (1983).

\bibitem{Klempt2021}
E.~Klempt and A.~V.~Sarantsev,
Singlet-octet-glueball mixing of scalar mesons,
Phys. Lett. B \textbf{826}, 136906 (2022).

\bibitem{Rosner1982}
J.~L.~Rosner,
Quark Content of Neutral Mesons,
Phys. Rev. D \textbf{27}, 1101 (1983).

\bibitem{tHooft1979}
G.~'t Hooft,
Naturalness, chiral symmetry, and spontaneous chiral symmetry breaking,
NATO Sci. Ser. B \textbf{59}, 135 (1980).

\bibitem{Ciambriello2024}
L.~Ciambriello, R.~Contino, A.~Luzio, M.~Romano and L.~X.~Xu,
A novel strategy to prove chiral symmetry breaking in QCD-like theories,
Phys. Lett. B \textbf{862}, 139367 (2025).

\bibitem{Parganlija2016}
D.~Parganlija and F.~Giacosa,
Excited Scalar and Pseudoscalar Mesons in the Extended Linear Sigma Model,
Eur. Phys. J. C \textbf{77},  450 (2017).


\bibitem{Csaki2023}
C.~Cs{\'a}ki, R.~Tito D'Agnolo, R.~S.~Gupta, E.~Kuflik, T.~S.~Roy and M.~Ruhdorfer,
On the dynamical origin of the $\eta^\prime$ potential and the axion mass,
JHEP \textbf{10}, 139 (2023).

\bibitem{Cuteri2021}
F.~Cuteri, O.~Philipsen and A.~Sciarra,
On the order of the QCD chiral phase transition for different numbers of quark flavours,
JHEP \textbf{11}, 141 (2021).

\bibitem{Dini2021}
L.~Dini, P.~Hegde, F.~Karsch, A.~Lahiri, C.~Schmidt and S.~Sharma,
Chiral phase transition in three-flavor QCD from lattice QCD,
Phys. Rev. D \textbf{105}, 3, 034510 (2022).

\bibitem{Pisarski2024}
R.~D.~Pisarski and F.~Rennecke,
Conjectures about the Chiral Phase Transition in QCD from Anomalous Multi-Instanton Interactions,
Phys. Rev. Lett. \textbf{132}, no.25, 251903 (2024).

\bibitem{Gribov}
V.~N.~Gribov,
Quantization of Nonabelian Gauge Theories,
Nucl. Phys. B \textbf{139}, 1,  (1978).

\bibitem{wett}
C. ~Wetterich, 
Spinors in euclidean field theory, complex structures and discrete symmetries,
Nucl. Phys. B \textbf{852}, 174 (2011).

\bibitem{ALVV22}
R.~L.~P.~G.~Amaral, V.~E.~R.~Lemes, O.~S.~Ventura and L.~C.~Q.~Vilar,
BRST view of spontaneous symmetry breaking,
Phys. Rev. D \textbf{105}, 12, 125007 (2022).

\end{thebibliography}
\end{document}